\documentclass{article}

\usepackage[utf8]{inputenc} 

\usepackage{amsmath, amssymb, amsfonts}
\usepackage{mathtools}
\usepackage{multirow}
\usepackage{wrapfig}
\usepackage{float} 
\usepackage{graphicx}
\usepackage{caption}
\usepackage{subcaption}
\usepackage{PRIMEarxiv}
\usepackage{xcolor}

\usepackage{algorithm}
\usepackage{algpseudocode}
\usepackage{tabularray}
\usepackage{ulem}

\usepackage[utf8]{inputenc} 
\usepackage[T1]{fontenc}    
\usepackage{hyperref}       
\usepackage{url}            
\usepackage{booktabs}       
\usepackage{amsfonts}       
\usepackage{nicefrac}       
\usepackage{microtype}      
\usepackage{lipsum}
\usepackage{fancyhdr}       
\usepackage{graphicx}       
\graphicspath{{media/}}     

\title{Discrepancy-Aware Contrastive Adaptation in Medical Time Series Analysis
}
\author{
  Yifan Wang\thanks{These authors contributed equally.} \\
  School of Medicine, The Chinese University of Hong Kong, Shenzhen \\
  \texttt{yifanwang13@link.cuhk.edu.cn} \\
  \And
  Hongfeng Ai\footnotemark[1] \\
  School of Medicine, The Chinese University of Hong Kong, Shenzhen \\
  \texttt{aihongfeng@cuhk.edu.cn} \\
  \And
  Ruiqi Li\footnotemark[1] \\
  University of the Chinese Academy of Sciences \\
  \texttt{liruiqi1@sia.cn} \\
  \And
  Maowei Jiang\footnotemark[1] \\
  Shenzhen International Graduate School, Tsinghua University \\
  \texttt{jiangmaowei@sia.cn} \\
  \And
  Ruiyuan Kang\\
  Wave and Machine Intelligence Department, \\
  Technology Innovation Institute \\
  \texttt{ruiyuan.kang@tii.ae} \\
  \And
  Jiahua Dong \\
  Mohamed bin Zayed University of \\
  Artificial Intelligence \\
  \texttt{dongjiahua1995@gmail.com} \\
  \And
  Cheng Jiang\thanks{Corresponding author.} \\
  School of Medicine, The Chinese University of Hong Kong, Shenzhen \\
  \texttt{jiangcheng@cuhk.edu.cn} \\
  \And
  Chenzhong Li\footnotemark[2] \\
  School of Medicine, The Chinese University of Hong Kong, Shenzhen \\
  \texttt{lichenzhong@cuhk.edu.cn} \\
}

\begin{document}
\maketitle

\begin{abstract}
In medical time series disease diagnosis, two key challenges are identified. First, the high annotation cost of medical data leads to overfitting in models trained on label-limited, single-center datasets. To address this, we propose incorporating external data from related tasks and leveraging AE-GAN to extract prior knowledge, providing valuable references for downstream tasks. Second, many existing studies employ contrastive learning to derive more generalized medical sequence representations for diagnostic tasks, usually relying on manually designed diverse positive and negative sample pairs. However, these approaches are complex, lack generalizability, and fail to adaptively capture disease-specific features across different conditions. To overcome this, we introduce LMCF (Learnable Multi-views Contrastive Framework), a framework that integrates a multi-head attention mechanism and adaptively learns representations from different views through inter-view and intra-view contrastive learning strategies. Additionally, the pre-trained AE-GAN is used to reconstruct discrepancies in the target data as disease probabilities, which are then integrated into the contrastive learning process. Experiments on three target datasets demonstrate that our method consistently outperforms other seven baselines, highlighting its significant impact on healthcare applications such as the diagnosis of myocardial infarction, Alzheimer’s disease, and Parkinson’s disease. We release the source code at xxxxx.
\end{abstract}


\section{Introduction}
Medical diagnosis tasks, such as Alzheimer's disease Parkinsonism require high precision and exhibit low tolerance for errors due to their critical impact on patient outcomes. In data-driven methods, time-series data is accordingly preferred to provide comprehensive information to support the accurate modeling \cite{rangapuram2018deep, gao2025auto}. However, the time-series data used are naturally complex with hierarchical architectures \cite{brunekreef2024kandinsky}. In contrast, the reliable labeled data is limited in its amount and diversity \cite{wen2020time, torres2021deep}, as they are often from the same center \cite{liu2021self, zhu2024mp}. These characteristics result in two key challenges: (1) difficulty in learning effective patterns and (2) increased risk of overfitting, which undermines the generalizability of data-driven medical diagnosis models.

To facilitate better pattern extraction, domain knowledge has been used to structure time-series data across multiple levels.\cite{wang2024contrast,brunekreef2024kandinsky,kiyasseh2021clocs}. In addition, representation learning methods, particularly contrastive learning, are employed to derive more informative representations. Although these endeavors improve the performance of ML model, the human bias introduced in data leveling and positive-negative pairing \cite{liu2023self,hong2021unbiased,hwang2022selecmix} in contrastive learning also constrain the model performance. On the other hand, to mitigate overfitting, external normal samples have been introduced as auxiliary resources to compensate for the limitations of training data and to alleviate the risk of “single-center bias"\cite{wu2022federated,harding201830}.
We admit the contribution of previous attempts, and think about that (1) whether we can design a more robust contrastive learning algorithm to avoid the labeling expense and bias, to better capture the effective patterns. (2) whether we can better utilize the external resource to avoid overfitting due to the data limitation. 

Accordingly, we propose \textbf{Discrepancy Adaptive Contrastive learning} for medical diagnosis tasks. It includes two modules:

(1) Discrepancy Estimator. Leveraging large-scale public normal data, we train a reconstructor based on an encoder-decoder architecture with GAN-style enhancement (AE-GAN). This reconstructor effectively models the distribution of normal samples. When encountering abnormal inputs, the resulting reconstruction error quantifies the discrepancy between the input and the learned normal distribution. This discrepancy signal supplements the limited abnormal data, provides additional supervision, and helps mitigate overfitting during training.

(2) Adaptive Contrastive Learner. Instead of relying on handcrafted positive-negative sample pairing, the learner employs Multi-Head Attention (MHA) to adaptively identify contrastive relationships. Specifically, it contrasts embeddings across attention heads (Inter-View) and across different patients (Intra-View). This mechanism enables the model to extract informative representations while reducing representational redundancy. The resulting features are further fine-tuned in a supervised manner for downstream medical diagnosis tasks.

The experimental results demonstrate that our method outperforms the SOTA methods across three representative clinical datasets, covering myocardial infarction, Alzheimer’s disease, and Parkinson’s disease. Notably, it still has decent performance when merely  10\% of the data is labeled, which highlights the effectiveness of the DAAC design.

\section{Related Work}
\label{sec:headings}
\paragraph{Medical time series}
Medical diagnosis often relies on physiological time-series signals such as EEG~\cite{savers1974mechanism}, ECG~\cite{lin2024ai}, EMG~\cite{ni2024survey}, and EOG~\cite{teng2020design}, which reflect complex temporal dynamics. Unlike images or texts, these data exhibit subtle and hierarchical variations~\cite{liu2023self}, increasing modeling difficulty. To address this, various representation learning methods have been proposed. TS2Vec~\cite{yue2022ts2vec} organizes data across temporal scales, CLOCS~\cite{kiyasseh2021clocs} models spatial-temporal invariance, and COMET~\cite{wang2024contrast} leverages four-level hierarchy to improve representation granularity. However, medical datasets are often limited to single-center collections with costly manual annotations~\cite{xu2022anomaly, zheng2023dabaclt}, leading to data scarcity and overfitting~\cite{luo2023infots, neumann2018discrepancy}.
\paragraph{Supervised learning for time series}
Supervised methods have shown strong performance in medical diagnostics, such as U-Sleep for EEG staging~\cite{perslev2021u}, L-SeqSleepNet for cross-night generalization~\cite{phan2023seqsleepnet}, and Inception-based networks for EOG classification~\cite{zeng2025multi}. To address annotation bottlenecks, techniques like SimGAN~\cite{golany2020simgans} generate realistic ECG signals, with GAN-based augmentation now mainstream for ECG~\cite{berger2023generative} and EEG~\cite{an2025imbalanced}. Yet, detecting rare events and adapting across domains remains challenging~\cite{chen2022cross, zhang2022self, jeon2023robust, chen2024addressing}, with supervised learning constrained by label imbalance and limited diversity.
\paragraph{Contrastive learning for time series}
Self-supervised learning—especially contrastive methods—has become a promising alternative for time-series modeling~\cite{zhang2024self}. By constructing positive and negative pairs through data augmentation~\cite{chen2020simple, yue2022ts2vec, eldele2023self, liu2024timesurl}, these approaches learn discriminative features without labels. For example, SoftCLT~\cite{lee2023soft} introduces soft assignment to improve flexibility but is sensitive to noise; CPC~\cite{oord2018representation} predicts future representations but suffers from negative sampling bias; TS-TCC~\cite{eldele2021time} contrasts temporal and contextual views yet struggles with dynamic or short sequences. Improving representation robustness in low-data regimes remains an open challenge.

\section{Proposed Method:DAAC}

\section{Proposed Method:DAAC}
As discussed above, in order to better extract more effective feature extraction and also avoid model overfitting on limited data, DAAC improved the paradigm of contrastive learning + downstream fine-tuning with two sophisticatedly designed modules. (1) \textbf{Discrepancy Estimator ($E_D$)} To avoid overfitting on limited data, we embed the information of the external normal data distribution into an Encoder-Decoder architecture reconstructor via GAN-style training (Sec.\ref{sec-discrepancy}), which is named as Discrepancy Estimator ($E_D$). Because the well-trained reconstructor captures the normal data distribution, and thus the reconstruction error reflects discrepancies of abnormal samples from normal ones. Appendix X demonstrates this experiment in detail. Such a design, utilizing the complementary information from normal data to alleviate the limitation of the given target data. The calculated discrepancy is used as a feature to augment the original data. (2) \textbf{Adaptive Contrastive Learner ($E_L$)}. To better capture effective features without further labeling (Sec. \ref{sec-adaptive}), we use the Multi-Head Attention (MHA) to embed features, which are sufficiently optimized through adaptively differentiating the embedding between attention heads and patients (inter-view and intra-view contrastive mechanisms, respectively). Meanwhile, features from diverse levels are used to provide more comprehensive information. 

\begin{figure*}[h]
\centering
\includegraphics[width=1\textwidth]{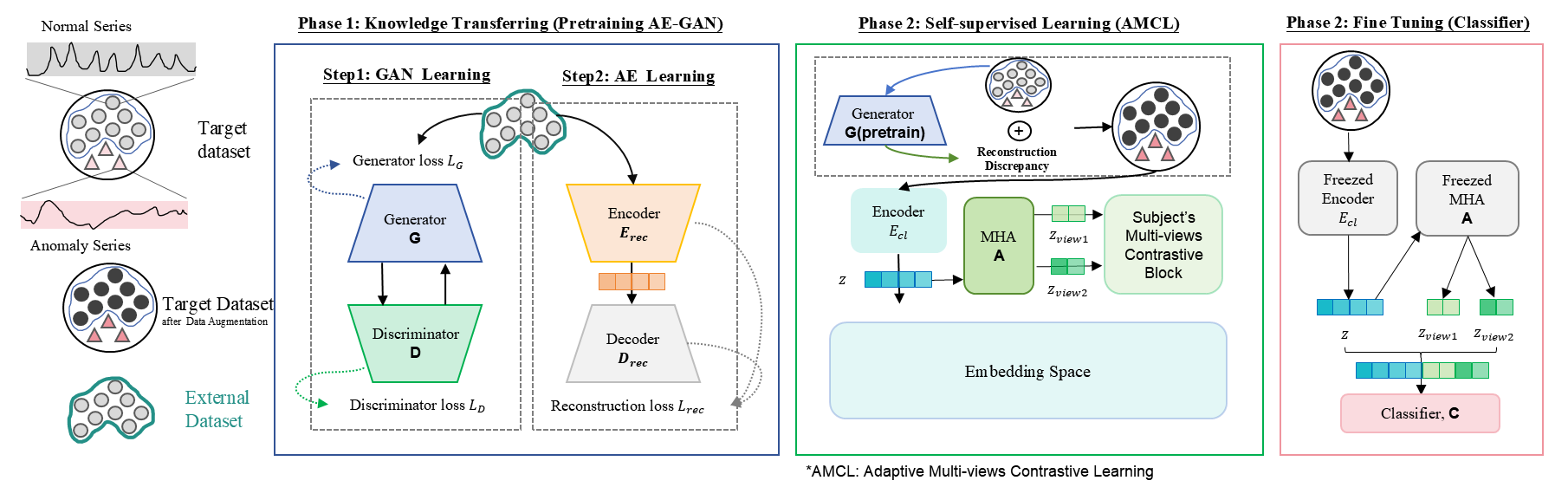}
\caption{Overview of our three-stage training framework. We first pretrain a Discrepancy Estimator ($E_D$) on normal data to extract reconstruction-based features. These features are then fused with raw inputs for contrastive representation learning with an MHA encoder. Finally, the learned encoder is fine-tuned with supervision for downstream tasks.}
\label{fig:train_proc}
\end{figure*}
\vspace{-10pt}

As shown in Figure \ref{fig:train_proc}, the training process consists of three stages. \textbf{Stage 1: Training and inference of Discrepancy Estimator}. A Discrepancy Estimator ($E_D$) is trained using data from normal patients in \( \mathcal{D}' \), whose samples are expressed as \( \mathbf{x}'_{i} \). Then, it is used to perform inference on all samples in \( \mathcal{D} \), whose samples are expressed as \( \mathbf{x}_{i} \), the calculated overall reconstruction error is used as an additional feature. fused with the raw data, resulting in a feature-augmented dataset with the number of features increased from \( F \) to \( F+1 \). \textbf{Stage 2: Training of Adaptive Contrastive Learner ($E_L$)} An encoder \( E_L \) with a multi-head attention block embed the augmented dataset into feature space, with the consideration of diverse data level and diverse views. Respecively, The well-trained encoder generates series-level features \( \mathbf{h}_{i} \in \mathbb{R}^{T \times C} \) with \( C \) dimensions, and view-level features \( \mathbf{g}_{i} \in \mathbb{R}^{T \times V \times d} \) with \( V \) views and \( d \) dimensions for each view. \textbf{Stage 3: Supervised fine-tuning} These representations are then used to train the classifier for specific downstream medical diagnostic tasks.

\subsection{Discrepancy Estimator}
\label{sec-discrepancy}

Although the target dataset \( \mathcal{D} \) contains limited and non-diverse data, the massive normal data  \( \mathcal{D}' \) from other institutions are available. Although it varies from  \( \mathcal{D} \), but both of them share similar characteristics, which accordingly provides complementary information for extracting better feature and alleviating overfitting on targeted dataset \( \mathcal{D} \). As mentioned above, we utilize the discrepancy to reflect the information of the normal dataset, an Encoder-Decoder Discrepancy Estimator ($E_D$ ) enhanced by GAN-style training is used to provide the function.
\subsubsection{Encoder-Decoder Generator}

The generator $G_{gen}$ is composed of an encoder $E_{gen}$ and a decoder $D_{gen}$, which work together to learn the distribution of healthy samples $\mathbf{x}'_i$ from the external dataset $D'$. In details, The encoder maps the healthy samples $\mathbf{x}'_i$ to a latent space representation and the decoder reconstructs the data from the representation to the generated data $\hat{\mathbf{x}}'_i$:

\vspace{-1.5em}
\begin{equation}
  \hat{\mathbf{x}}'_i = G_{gen}(\mathbf{x}'_i) = D_{gen}(E_{gen}(\mathbf{x}'_i))
\end{equation}
\vspace{-2em}

The generator is trained to minimize the reconstruction loss, ensuring that the generated data $\hat{\mathbf{x}}'_i$ closely approximates the original healthy samples $\mathbf{x}'_i$. This enables the generator to capture the underlying distribution of healthy data, which is critical for subsequent feature augmentation and knowledge transfer.

\subsubsection{Discriminator}

In order to maximally enhance the performance of generator, The discriminator $D_{dis}$ is used to distinguish between real healthy samples $\mathbf{x}'_i$ from the external dataset $D'$ and generated data $\hat{\mathbf{x}}'_i$. The output of $D_{dis}$ represents the probability that the sample is from healthy subject. The details about the loss function can be seen in Appendix~\ref{sec:appendix_d}.
\subsubsection{Inference on Internal Data}
After pretraining, we use the Generator model to perform inference on all samples $\mathbf{\bar x}_i$ of  $D$, and the reconstruction error $\mathcal{E}$ is computed by:

\vspace{-1em}
\begin{equation}
\mathcal{E} = \text{MSE}(G_{gen}(\mathbf{\bar x}_i), \mathbf{\bar x}_i)
\end{equation}
\vspace{-1.5em}

Since the generator is pretrained on healthy samples $\mathbf{x}'_i$ from $D'$, the reconstruction error $\mathcal{E}$ is smaller for healthy samples $\mathbf{x}_i$ in the target dataset and larger for abnormal samples $\mathbf{x}_i$. Thus, the magnitude of $\mathcal{E}$ indirectly reflects the abnormality of the sample $\mathbf{x}_i$, i.e., the probability of disease. In our design, we compute sequence-level reconstruction error rather than point-wise error. This decision is motivated by the observation that point-wise errors are often highly sensitive to local fluctuations and noise, which are common in medical time series due to physiological variability or sensor artifacts. In contrast, sequence-level feedback captures the global consistency and overall structural deviation from expected normal patterns, leading to more robust and stable anomaly signals. Additional details can be found in Appendix~\ref{de_study}
   
The reconstruction error $\mathcal{E}$ is concatenated with the original features $x_i$ of the target dataset to form an augmented feature set:

\vspace{-1.5em}
\begin{equation}
\mathbf{x}_i = \text{concat}(\mathbf{\bar x}_i,\mathcal{E})
\end{equation}

The augmented dataset $ {\mathbf{x}_i} $ is used for subsequent model development.

\subsection{Adaptive Contrastive Learner}
\label{sec-adaptive}
Conventional contrastive learning relies on manually designed heuristics for constructing positive and negative pairs, which may introduce noise. To address this limitation, we propose an adaptive contrastive loss that leverages multi-head attention to guide the model in learning which views should be pulled together or pushed apart. This approach eliminates the reliance on external labels and handcrafted pairing strategies, while producing robust and semantically rich representations. The Adaptive Contrastive Learner $E_L$ of our approach is designed with two components.

The first component is a dilated convolutional network $h(\mathbf{x})$, which broadens the receptive field of the sequence. This network is crucial for capturing a broader range of information and enhancing the generalizability of the temporal representations. On the other hand, inspired from the local transformations to craft diverse data views in the latent space for contrastive learning \cite{schneider2022detecting}, we recognized that MHA can significantly reduce the manual cost of designing positive and negative sample pairs from different perspectives of data. Consequently, we incorporated a second component into our encoder, which is our innovative learnable multi-views network $g(\mathbf{x})$.

\begin{wrapfigure}{r}{0.25\textwidth}
  \centering
  \includegraphics[width=0.25\textwidth]{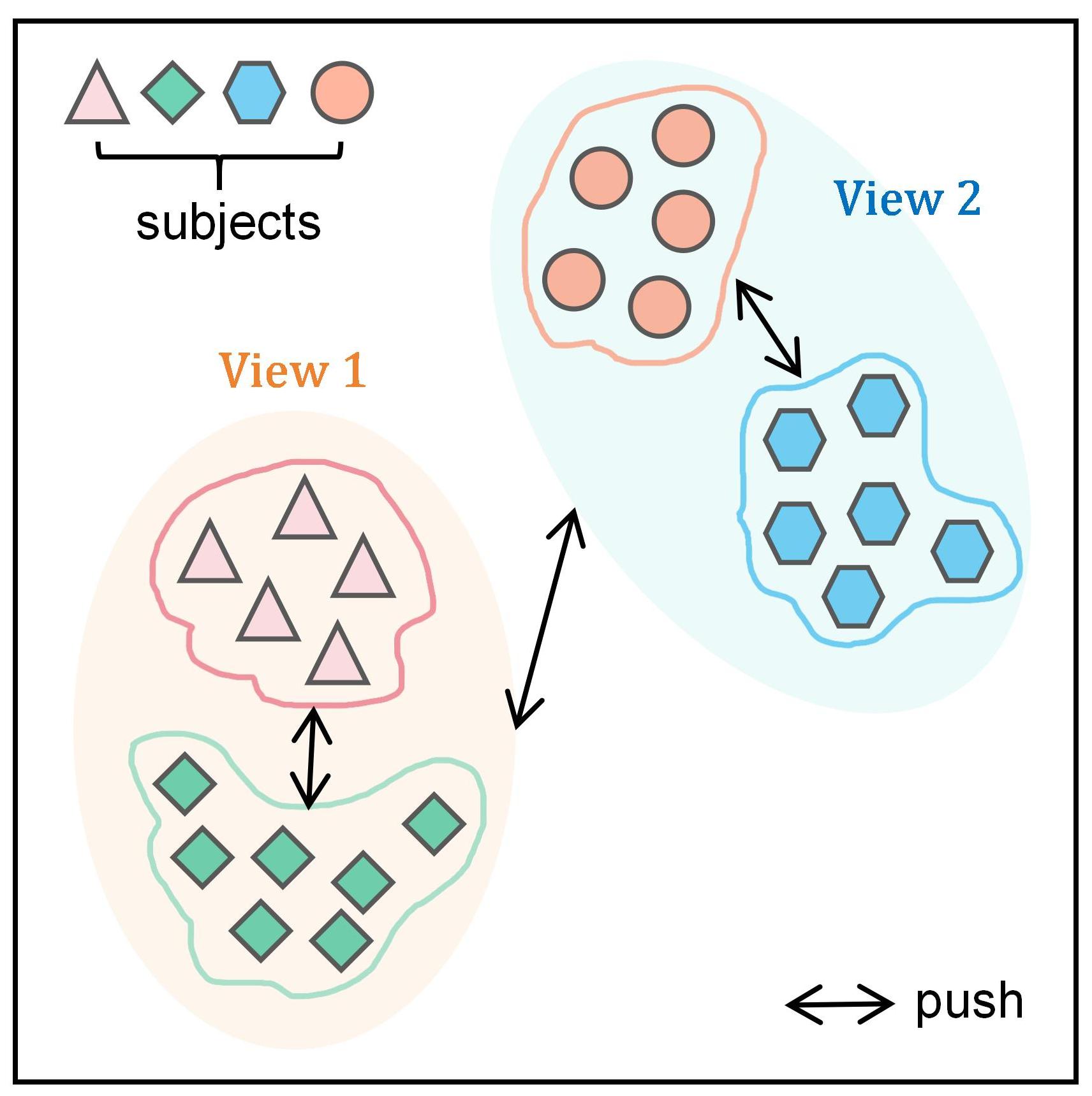}
  \caption{Mechanism of Multi-View Contrastive Learning}
  \label{fig:view_cl}
\end{wrapfigure}

It introduces MHA to contrastive learning for extracting varying view representations adaptively. By leveraging both inter-wise and intra-wise contrastive losses, we ensure that the representations not only encompass diverse views but also exhibit distinct separability among different subjects within each view as Figure \ref{fig:view_cl} shown.

Furthermore, to capture the intrinsic hierarchical information embedded in medical data, we adopt the concept from COMET \cite{wang2024contrast}. Our contrastive learning considers subject, trial, epoch, and temporal, and optimizes the model using multiple InfoNCE losses \cite{oord2018representation}. Ultimately, we obtain representations that are beneficial for downstream medical time-series tasks.



\subsubsection{Subject-wise Contrastive Loss $L_{S}$} 
Subject may exhibit unique physiological characteristics and pathological variations. Consequently, the subject-wise contrastive learning is designed to enhance the discriminative representation of subject-specific features by increasing the similarity among samples from the same subject and the separability among samples from different subjects.

Specifically, let \( \mathbf{x}_{i,s} \) denote the \( i \)-th anchor sample in the given batch, which is from subject \( s \). The positive sample set \( S_{i}^{+} \) for \( \mathbf{x}_{i,s} \) consists of all other samples \( \mathbf{x}_{j, s^{+}} \) with the same subject \( s^{+} \). Correspondingly, the negative sample set \( S_{i}^{-} \) includes all samples \( \mathbf{x}_{j, s^{-}} \) from different subjects \( s^{-} \). Based on this delineation of positive and negative samples, we define the subject-level contrastive loss function as follows:

\vspace{-1.5em}
\begin{equation}
\small
L_{S} = \frac{-1}{M}\sum_{i=1}^{M} \sum_{\mathbf{x}_{j,s^{+}} \in S_{i}^{+}} \log \frac{ \exp\left(s(\mathbf{h}_{i,s}, \mathbf{h}_{j,s^{+}}) / \tau\right)}{\sum_{\mathbf{x}_{j,s^{-}} \in S_{i}^{-}} \exp\left({s(\mathbf{h}_{i,s}, \mathbf{h}_{j,s^{-}}) / \tau}\right)} 
\end{equation}
\vspace{-1.5em}

where \( j \neq i \) and \( s(\cdot, \cdot) \) denotes the cosine similarity between the respective representations. $\exp(\cdot)$ is the exponential function and \( M \) represents the batch size. \( \mathbf{h}_{i,s} \) corresponds to the output of the network \( h(x) \) applied to the sample \( \mathbf{x}_{i,s} \), i.e., \( \mathbf{h}_{i,s} = h(\mathbf{x}_{i,s}) \). Similarly, \( \mathbf{h}_{j,s^{+}} \) and \( \mathbf{h}_{j,s^{-}} \) are derived from \( h(\mathbf{x}_{j,s^{+}}) \) and \( h(\mathbf{x}_{j,s^{-}}) \). The parameter \( \tau \) serves as a temperature parameter. 


\subsubsection{Trial-wise Contrastive Loss $L_{R}$}

Trials may demonstrate distinctive experimental conditions and variations in outcomes. To address this, trial-wise contrastive learning enhances trial-specific feature representation by promoting similarity within trials and separability between trials.

Concretely, assume \( \mathbf{x}_{i,r} \) represent the \( i \)-th anchor sample in the batch, sourced from trial \( r \). The set of positive samples \( R_{i}^{+} \) associated with \( \mathbf{x}_{i,r} \) encompasses all other samples \( \mathbf{x}_{j, r^{+}} \) that are from the same trial \( r^{+} \). Conversely, the set of negative samples \( R_{i}^{-} \) consists of all samples \( \mathbf{x}_{j, r^{-}} \) originating from different trials \( r^{-} \). Therefore, we establish the trial-wise contrastive loss function as below:
\vspace{-0.5em}
\begin{equation}
\small
L_{R} = \frac{-1}{M} \sum_{i=1}^{M} \sum_{\mathbf{x}_{j,r^{+}} \in R_{i}^{+}} \log \frac{ \exp\left(s(\mathbf{h}_{i,r}, \mathbf{h}_{j,r^{+}}) / \tau\right)}{\sum_{\mathbf{x}_{j,r^{-}} \in R_{i}^{-}} \exp\left({s(\mathbf{h}_{i,r}, \mathbf{h}_{j,r^{-}}) / \tau}\right)} 
\end{equation}
\vspace{-1.5em}

where, \( \mathbf{h}_{i,r} \) denotes the representation obtained by applying the network \( h(\mathbf{x}) \) to the sample \( \mathbf{x}_{i,r} \), formally expressed as \( \mathbf{h}_{i,r} = h(\mathbf{x}_{i,r}) \). Consistent with this notation, \( \mathbf{h}_{j,r^{+}} \) and \( \mathbf{h}_{j,r^{-}} \) represent the feature representations resulting from the application of \( h(\mathbf{x}) \) to the corresponding positive and negative samples \( \mathbf{x}_{j,r^{+}} \) and \( \mathbf{x}_{j,r^{-}} \).

\subsubsection{Epoch-wise Contrastive Loss $L_{E}$}

For epoch-wise contrastive learning, our hypothesis is that a sample which undergoes minor modifications, like temporal masking, should maintain a similar pattern compared to other sample. Hence, for each sample \( \mathbf{x}_i \) in batch, we employ a data augmentation strategy that is random continuous masking. This strategy involves applying two independent augmentations to \( \mathbf{x}_i \), resulting in augmented samples \( \tilde{\mathbf{x}}_{i}^{1} \) and \( \tilde{\mathbf{x}}_{i}^{2} \). Since these augmented samples originate from the same anchor sample \( \mathbf{x}_{i} \), they are encouraged to have high similarity between their feature representations and are considered as positive pair $(\tilde{\mathbf{x}}_{i}^{1}, \tilde{\mathbf{x}}_{i}^{2})$. In contrast, negative sample pairs are formed from augmented samples derived from different anchor samples, such as $\left ( \tilde{\mathbf{x}}_{i}^{1}, \tilde{\mathbf{x}}_{j}^{1} \right )$ and $\left ( \tilde{\mathbf{x}}_{i}^{1}, \tilde{\mathbf{x}}_{j}^{2} \right )$. Finally, the epoch-level loss function is formulated as:
\begin{equation}
\small
L_{E}=\frac{-1}{M}\sum_{i=1}^{M}\log\frac{\exp\left(\tilde{\mathbf{h}}_{i}^{1}\cdot \tilde{\mathbf{h}}_{i}^{2}\right)}{\sum\limits_{j=1,j\ne i}^{M}\left ( \exp\left ( \tilde{\mathbf{h}}_{i}^{1}\cdot \tilde{\mathbf{h}}_{j}^{1} \right )  +\exp\left (\tilde{\mathbf{h}}_{i}^{1}\cdot \tilde{\mathbf{h}}_{j}^{2}\right ) \right ) } 
\end{equation}

where $\tilde{\mathbf{h}}_{i}^{1}=h(\tilde{\mathbf{x}}_{i}^{1})$ and $\tilde{\mathbf{h}}_{i}^{2}=h(\tilde{\mathbf{x}}_{i}^{2})$.

\subsubsection{Temporal-wise Contrastive Loss $L_{T}$}

The objective of this loss function is to ensure that two augmented samples originating from the same sample exhibit similar representations when observed at the same timestamp, while their representations differ when observed at other timestamps. Temporal-wise contrastive learning leverages the temporal consistency within the data. In this context, let us consider the anchor sample \(\mathbf{x}_{i,t}\) of the \(i\)-th data point at time \(t\) within a batch. By applying temporal masking, we generate two augmented samples as \(\tilde{\mathbf{x}}_{i,t}^{1}\) and \(\tilde{\mathbf{x}}_{i,t}^{2}\). The pair \((\tilde{\mathbf{x}}_{i,t}^{1}, \tilde{\mathbf{x}}_{i,t}^{2})\) is classified as a positive sample pair because they are derived from the same anchor sample at timestamp \(t\). Conversely, negative sample pairs are defined as \((\tilde{\mathbf{x}}_{i,t}^{1}, \tilde{\mathbf{x}}_{i,t^{-}}^{1})\) and \((\tilde{\mathbf{x}}_{i,t}^{1}, \tilde{\mathbf{x}}_{i,t^{-}}^{2})\), where \(t \neq t^{-}\). These negative pairs are constructed from augmented samples originating from different temporal points, thereby encouraging the model to learn distinct representations for different time observations. The Temporal-wise contrastive loss is shown as:
\small
\begin{align}
L_{T} &= \frac{-1}{T\cdot M}\sum_{t=1}^{T }\sum_{i=1}^{M}\log Q_{i,t} \\
Q_{i,t} &= 
\frac{\exp\left(\tilde{\mathbf{h}}_{i,t}^{1}\cdot \tilde{\mathbf{h}}_{i,t}^{2}\right)}
{\sum\limits_{t=1,t\ne t^{-}}^{T}
\left( 
\exp\left ( \tilde{\mathbf{h}}_{i,t}^{1}\cdot \tilde{\mathbf{h}}_{i,t^{-}}^{1} \right )  
+\exp\left (\tilde{\mathbf{h}}_{i,t}^{1}\cdot \tilde{\mathbf{h}}_{i,t^{-}}^{2}\right ) 
\right)}
\end{align}

where $\tilde{\mathbf{h}}_{i,t}^{1}=h(\tilde{\mathbf{x}}_{i,t}^{1})$. Similarly, analogous notations with $h$ are derived from the network $h(\mathbf{x})$.

\subsubsection{Inter-view Contrastive Loss $L_{IRV}$} The inter-view contrastive loss we propose aims to encourage the model to capture informative features across different augmented views adaptively. Essentially, inter-view loss drives the representations from different views to be distinguishable from each other, thereby promoting the diversity of views.

Let us consider a batch of samples as $\mathbf{x}_i$, where $i = 1, \ldots, M$. For each sample $\mathbf{x}_i$, we apply data augmentation to generate two augmented samples, $\tilde{\mathbf{x}}_{i}^{1}$ and $\tilde{\mathbf{x}}_{i}^{2}$. These augmented samples are subsequently fed into the MHA module, denoted as $g(\mathbf{x})$. This module generates representations $\tilde{\mathbf{g}}_{i,v}^{k}$ for each augmented branch $k \in \left \{ 1,2 \right \}$ and different views $v$, where $g(\mathbf{x}_{i}^{k}) = \tilde{\mathbf{g}}_{(i,v)}^{k}$. The view representations from the given batch are then aggregated into $\tilde{\mathbf{g}}_{v}^{k} = \left \{ \tilde{\mathbf{g}}_{1,v}^{k}, \tilde{\mathbf{g}}_{2, v}^{k}, \ldots, \tilde{\mathbf{g}}_{M, v}^{k}\right \} $.

Assume we have two views for each branch that is $V=2$, To facilitate inter-view contrastive learning, we define positive sample pairs as those from the same view, such as $(\tilde{\mathbf{g}}_{1}^{1}, \tilde{\mathbf{g}}_{1}^{2})$ and $(\tilde{\mathbf{g}}_{2}^{1}, \tilde{\mathbf{g}}_{2}^{2})$. Conversely, negative sample pairs are those from different views, such as $(\tilde{\mathbf{g}}_{1}^{1}, \tilde{\mathbf{g}}_{2}^{1})$ and $(\tilde{\mathbf{g}}_{2}^{1}, \mathbf{g}_{1}^{2})$. Hence, the inter-view contrastive loss is formulated as:

\begin{equation}
L_{IRV} = \frac{-1}{V}\sum_{i=1}^{V}  \log \frac{\exp(\text{s}(\tilde{\mathbf{g}}_{i}^{1}, \tilde{\mathbf{g}}_{i}^{2}) )}{\sum_{j=1,j\ne i}^{V} \exp(\text{s}(\tilde{\mathbf{g}}_{i}^{1}, \tilde{\mathbf{g}}_{j}^{2}) )} 
\end{equation}

\subsubsection{Intra-View Contrastive Loss $L_{IAV}$} In addition to inter-view loss, we also introduce the intra-view loss to further refine the representations within the same view. This loss is designed to enhance the discriminative power of representations by contrasting samples from the same view but different subjects, thus encouraging the model to capture subject-specific view features more effectively.

Similarly, assume we have subject-specific samples $\mathbf{x}_{i,s}$, where $i$ ranges from $1$ to $M$. For each of them, we apply data augmentation, resulting in augmented samples $\tilde{\mathbf{x}}_{i}^{1}$ and $\tilde{\mathbf{x}}_{i}^{2}$. These augmented samples are then processed through a MHA network $g(\mathbf{x})$, which produces subject-specific view representations $\tilde{\mathbf{g}}_{i,v}^{k}$ for each augmented branch $k \in \left \{ 1,2 \right \} $ and various views $v$, i.e., $\tilde{\mathbf{g}}_{i,v,s}^{k} = g(\mathbf{x}_{i,s}^{k})$.

Suppose we have $V$ views and $S$ subjects. To enable intra-view contrastive learning, we identify positive sample pairs as those from the same view and subject. Therefore, we have positive pairs that are $\left (  \tilde{\mathbf{g}}_{i,v,s}^{1}, \tilde{\mathbf{g}}_{i,v^{+},s^{+}}^{2}\right ) $. On the other hand, The negative sample pairs are defined as the samples that differ in subject at the given view, which are $\left (  \tilde{\mathbf{g}}_{i,v,s}^{1}, \tilde{\mathbf{g}}_{i,v^{+},s^{-}}^{2}\right ) $

\begin{equation}
\begin{aligned}
L_{IAV} &= \frac{-1}{V\cdot M}\sum_{v=1}^{V}\sum_{i=1}^{M}
\sum_{\mathbf{x}_{j,v^{+},s^{+}} \in S_{i}^{+}} 
\log K_{i,v,s} \\
K_{i,v,s} &= \frac{\exp({s( \tilde{\mathbf{g}}_{i,v,s}^{1} \cdot \tilde{\mathbf{g}}_{j,v^{+},s^{+}}^{2})/\tau})}
{\sum\limits_{\mathbf{x}_{j,v^{a+},s^{-}} \in S_{i}^{-}}  
\exp({s(\tilde{\mathbf{g}}_{i,v,s}^{1} \cdot \tilde{\mathbf{g}}_{j,v^{+},s^{-}}^{2})/\tau })}
\end{aligned}
\end{equation}

where $S_{i}^{+}$ includes the augmented sample set of $\tilde{\mathbf{x}}_{j,v^{+},s^{+}}^{2}$ that belong to the same subject as the given augmented sample $\tilde{\mathbf{x}}_{i,v,s}^{2}$ under the same view $v$. Conversely, $S_{i}^{-}$ consists of the augmented samples that come from the same view but different subjects.

Thus, our view loss comprises both inter-view and intra-view contrastive losses:

\vspace{-1.5em}
\begin{equation}
\mathcal{L}_{V} = L_{IRV} + L_{IAV}
\end{equation}
\vspace{-1.5em}

Finally, the overall contrastive loss function $L$ is defined as:

\vspace{-1.5em} 
\begin{equation}
\mathcal{L} = \lambda_{S} \cdot L_{S} + \lambda_{R} \cdot L_{R} + \lambda_{E} \cdot L_{E} + \lambda_{T} \cdot L_{T} + \lambda_{V} \cdot L_{V}
\end{equation}
\vspace{-1.5em} 

where $\lambda_{S}:\lambda_{R}:\lambda_{E}:\lambda_{T}:\lambda_{V}=1:1:1:1:2$ in our practice. The details about loss weight sensitivity could be seen in Appendix \ref{Loss Weight Sensitivity}

\subsection{Fine Tuninig}
In order to use the Encoder $E_L$ of DAAC for downstream tasks, we utilize fine-tuning as follows:

\vspace{-1.5em}
\begin{align}
\hat{y}_{i} = C(E_L\left (  \mathbf{x}_{i} \right ) ) = C(h(\mathbf{x}_{i}), g(\mathbf{x}_{i}))
\end{align}
\vspace{-1.5em}

where $\hat{y}$ is the classification result for downstream medical disease diagnosis, and the training optimization objective is binary cross-entropy.

\section{Experiments}
This study utilized three target datasets corresponding to Alzheimer’s disease, myocardial infarction, and Parkinson’s disease. Additionally, three external datasets from independent institutions were incorporated to facilitate cross-center knowledge transfer. A detailed description of all datasets and the preprocessing protocols is provided in Appendix \ref{sec:appendix_c}.  All datasets are publicly available and had received institutional review board (IRB) approval prior to their release. 

All experiments were conducted on a workstation equipped with four NVIDIA RTX 4090 GPUs (24GB each). The models were implemented in PyTorch 1.13.1 with CUDA 11.6. All results were obtained under the same hardware configuration to ensure fairness and consistency in comparison.

\subsection{Results on Partial Fine-tuning}
To evaluate the representation quality of our proposed DAAC quickly, we adopt the Partial Fine-Tuning (PFT). PFT refers to combining the Encoder with frozen parameters and a trainable logistic regression linear classifier $C$. In this setup, we utilizes 100\% of the labeled data for training the supervised classification. The PFT results on the AD dataset are shown in Table~\ref{tab:Partial Fine-tuning Results}. Our proposed DAAC model excels in all metrics, including Accuracy, Precision, Recall, F1 score, AUROC, and AUPRC, demonstrating the power of our representation learning.
\vspace{-5pt}
\begin{table}[H]
\centering
\scriptsize 
\caption{Partial Fine-tuning Results. 
\textbf{ACL} = Adaptive Contrastive Learner; 
\textbf{DE} = Discrepancy Estimator; 
\textbf{COMET+ACL}, \textbf{COMET+DE} represent ablations with only ACL or DE added to COMET; 
\textbf{DAAC} is the final model combining both ACL and DE. Bold indicates the best performance, and \underline{underline} denotes the second best.}
\label{tab:Partial Fine-tuning Results}
\resizebox{\textwidth}{!}{
\begin{tabular}{l|c|c|c|c|c|c}
\hline
\textbf{Models} & \textbf{Accuracy} & \textbf{Precision} & \textbf{Recall} & \textbf{F1 score} & \textbf{AUROC} & \textbf{AUPRC} \\ \hline
\textbf{TS2vec} & $66.48\pm5.33$ & $67.72\pm5.09$ & $67.40\pm5.20$ & $66.32\pm5.46$ & $74.12\pm6.88$ & $72.96\pm7.21$ \\
\textbf{TF-C} & $77.03\pm2.80$ & $75.79\pm5.07$ & $64.27\pm5.03$ & $64.85\pm5.56$ & $80.71\pm4.03$ & $79.27\pm4.15$ \\
\textbf{Mixing-up} & $46.16\pm1.38$ & $52.62\pm4.90$ & $50.81\pm1.32$ & $37.37\pm1.98$ & $64.42\pm6.49$ & $62.85\pm6.07$ \\
\textbf{TS-TCC} & $59.71\pm8.63$ & $61.66\pm8.63$ & $60.33\pm3.26$ & $58.66\pm8.39$ & $67.53\pm10.04$ & $68.33\pm9.37$ \\
\textbf{SimCLR} & $57.16\pm2.05$ & $56.67\pm3.91$ & $53.57\pm2.21$ & $49.11\pm4.26$ & $56.67\pm3.91$ & $52.10\pm1.41$ \\
\textbf{CLOCS} & $66.99\pm2.84$ & $67.17\pm2.96$ & $67.33\pm2.99$ & $66.91\pm2.88$ & $73.71\pm3.62$ & $72.58\pm3.54$ \\
\textbf{COMET} & $76.09\pm4.21$ & $77.36\pm3.97$ & $74.68\pm4.62$ & $74.80\pm4.83$ & $81.30\pm4.97$ & $80.50\pm5.31$ \\ \hline
\textbf{COMET+ACL} & $77.42\pm4.40$ & $78.64\pm3.73$ & $76.11\pm4.98$ & $76.26\pm5.26$ & $82.44\pm5.19$ & $81.30\pm5.56$ \\
\textbf{COMET+DE} & $\underline{79.60\pm6.01}$ & $\textbf{80.41}\pm\textbf{5.86}$ & $\underline{78.50\pm6.51}$ & $\underline{78.75\pm6.46}$ & $\textbf{86.32}\pm\textbf{6.54}$ & $\textbf{85.85}\pm\textbf{6.99}$ \\
\textbf{DAAC} & $\textbf{80.48}\pm\textbf{5.97}$ & $\underline{80.34\pm6.00}$ & $\textbf{80.04}\pm\textbf{6.27}$ & $\textbf{80.12}\pm\textbf{6.22}$ & $\underline{85.96\pm7.17}$ & $\underline{85.40\pm7.52}$ \\
\hline

\end{tabular}
}
\end{table}

\vspace{-15pt}
\subsection{Results on Full Fine-tuning}
Unlike PFT, Full Fine-tuning (FFT) adapts all model parameters to the given target medical task, which means the Encoder and the classifier are trained together with supervision. FFT further enhances the representation in the downstream task.
The results of FFT on the target dataset are exemplified using the AD, PD, PTB dataset in Table ~\ref{tab:performance_comparison}.
\subsection{Visualization of Views}
To verify the learnable multi-view learning capability of ACL, we employed the UMAP \cite{mcinnes2018umap} to reduce the dimensionality of view representations to two dimensions and visualize them as shown in Fig. \ref{fig:view_representation}. The visualizations provide strong validation from the following two aspects:

\begin{figure*}[t]
\centering
\includegraphics[width=1\textwidth]{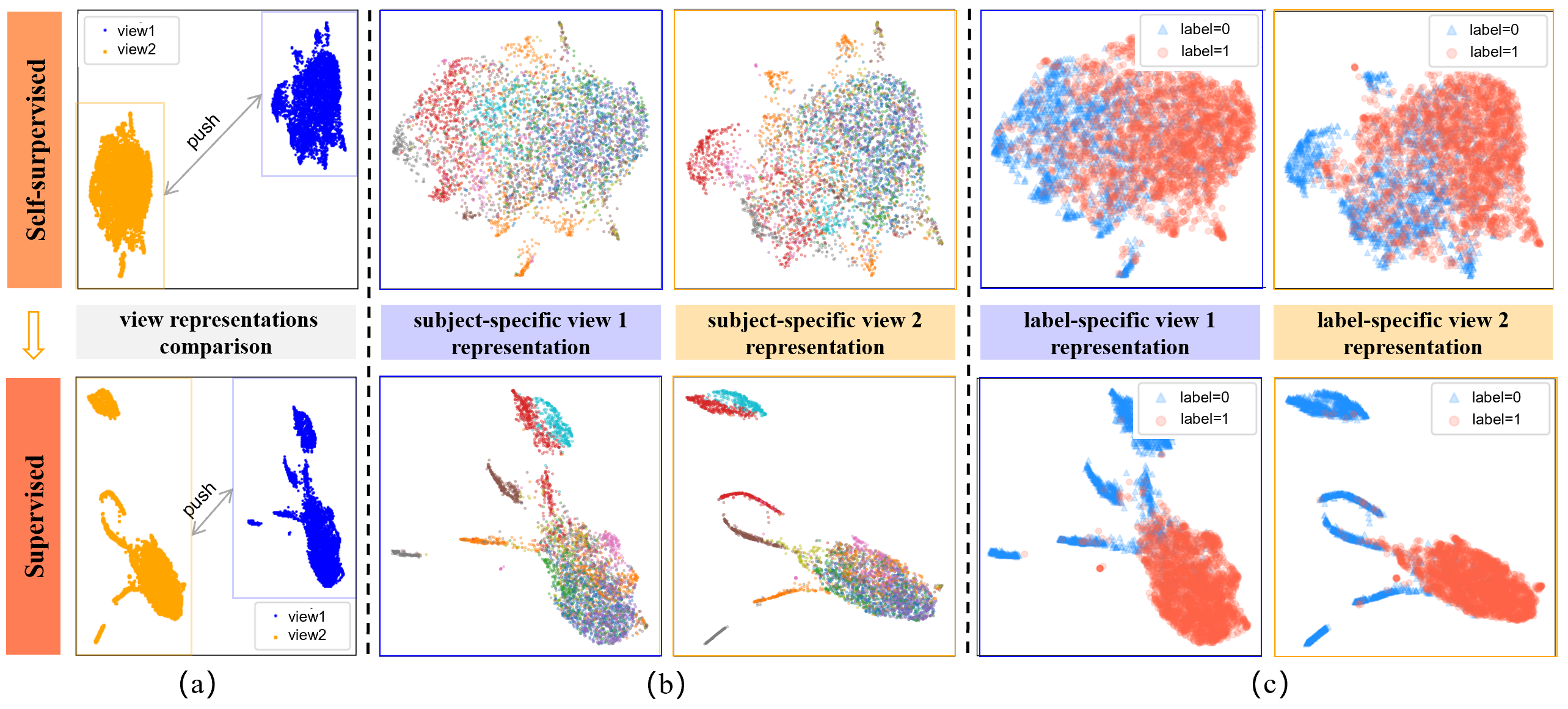}
\caption{Visualization of Multi-View Representations on the AD Dataset. The figure compares contrastive view representations obtained through self-supervised contrastive learning and 100\% supervised fine-tuning.}
\vspace{-20pt}
\label{fig:view_representation}
\end{figure*}
\textbf{The Disparities between Diverse Views}: By comparing the view representations as Fig. \ref{fig:view_representation} (a), it can be seen that the designed inter-view contrastive loss enables the representations of view 1 and view 2 under different samples to show obvious disparities, effectively distinguishing the representations of each view.

\textbf{The Intra-view Subject Discrepancies}: The intra-view contrastive loss we propose fully considers the subject identity, setting the positive samples as those samples with the same view and subject as the anchor sample, and the negative samples as those with the same view but different subjects with the anchor sample. From the visualization results of patient-specific view representations in Fig. \ref{fig:view_representation} (b), it can be found that there are differences in the representations of some different subjects under each view. In the self-supervised part, although label is not introduced, when observing the representations of different views with label coloring in Fig. \ref{fig:view_representation} (c), the representation differences between diseased subjects and normal subjects in the AD dataset can still be found. This indicates that the view representations have good generalization ability, which is beneficial for downstream disease diagnosis tasks.

In addition, Fig. \ref{fig:view_representation} also shows that after adding 100\% labeled data for supervised fine-tuning, the clustering of view representations is further enhanced. Overall, through this visualization, it is sufficient to prove the effectiveness of  the adaptive contrastive learning.

\begin{table}[htb]
\centering
\caption{Full fine-tuning results of AD, TDBrain, PTB datasets. We use 100\% and 10\% of labeled data for fine-tuning. Model names are consistent with Table~\ref{tab:Partial Fine-tuning Results}. Bold indicates the best performance, and \underline{underline} denotes the second best.}
\label{tab:performance_comparison}
\resizebox{\textwidth}{!}{
\begin{tabular}{c|clllllll}
\hline
\textbf{Datasets} & \textbf{Fraction} & \textbf{Models} & \textbf{Accuracy} & \textbf{Precision} & \textbf{Recall} & \textbf{F1 score} & \textbf{AUROC} & \textbf{AUPRC} \\ \hline
\multirow{20}{*}{\textbf{AD}} & \multirow{10}{*}{\textbf{100\%}} & \textbf{TS2vec} & $81.26 \pm 2.08$ & $81.21 \pm 2.14$ & $81.34 \pm 2.04$ & $81.12 \pm 2.06$ & $89.20 \pm 1.76$ & $88.94 \pm 1.85$ \\
 &  & \textbf{TF-C} & $75.31 \pm 8.27$ & $75.87 \pm 8.73$ & $74.83 \pm 8.98$ & $74.54 \pm 8.85$ & $79.45 \pm 10.23$ & $79.33 \pm 10.57$ \\ 
 &  & \textbf{Mixing-up} & $65.68 \pm 7.89$ & $72.61 \pm 4.21$ & $68.25 \pm 6.97$ & $63.98 \pm 9.92$ & $84.63 \pm 5.04$ & $83.46 \pm 5.48$ \\ 
 &  & \textbf{TS-TCC} & $73.55 \pm 10.00$ & $77.22 \pm 6.13$ & $73.83 \pm 9.65$ & $71.86 \pm 11.59$ & $86.17 \pm 5.11$ & $85.73 \pm 5.11$ \\ 
 &  & \textbf{SimCLR} & $54.77 \pm 1.97$ & $50.15 \pm 7.02$ & $50.58 \pm 1.92$ & $43.18 \pm 4.27$ & $50.15 \pm 7.02$ & $50.42 \pm 1.06$ \\ 
 &  & \textbf{CLOCS} & $78.37 \pm 6.00$ & $83.99 \pm 2.11$ & $76.14 \pm 7.03$ & $75.78 \pm 7.93$ & $91.17 \pm 2.51$ & $90.72 \pm 3.05$ \\ 
 &  & \textbf{COMET} & $84.50 \pm 4.46$ & $88.31 \pm 2.42$ & $82.95 \pm 5.39$ & $83.33 \pm 5.15$ & $94.44 \pm 2.37$ & $94.43 \pm 2.48$ \\ 
&  & \textbf{COMET+DE} & $ 84.82\pm 10.43$ & $88.49 \pm 5.71$ & $83.42 \pm 11.99$ & $83.08 \pm 12.99$ & $93.97 \pm 4.97$ & $93.71 \pm 5.25$ \\ 
&  & \textbf{COMET+ACL} & $ \underline{91.16\pm 4.14}$ & $\underline{92.10\pm3.66}$ & $\underline{90.50 \pm4.51}$ & $\underline{90.88 \pm 4.33}$ & $\underline{96.22 \pm 2.76}$ & $\underline{96.03 \pm 2.82}$ \\ 
&  & \textbf{DAAC} & $ \textbf{93.23}\pm \textbf{5.25}$ & $\textbf{94.01} \pm \textbf{3.98}$ & $\textbf{92.71} \pm \textbf{5.86}$ & $\textbf{92.97} \pm \textbf{5.65}$ & $\textbf{98.03} \pm \textbf{1.71}$ & $\textbf{98.03} \pm \textbf{1.75}$ \\ 
  \cline{2-9} 
 & \multirow{10}{*}{\textbf{10\%}} & \textbf{TS2vec} & $73.28 \pm 4.34$ & $74.14 \pm 4.33$ & $73.52 \pm 3.77$ & $73.00 \pm 4.18$ & $81.66 \pm 5.20$ & $81.58 \pm 5.11$ \\ 
 &  & \textbf{TF-C} & $75.66 \pm 11.21$ & $75.48 \pm 11.48$ & $75.58 \pm 11.59$ & $75.38 \pm 11.47$ & $81.38 \pm 14.19$ & $81.56 \pm 13.68$ \\
 &  & \textbf{Mixing-up} & $59.38 \pm 3.33$ & $64.85 \pm 4.38$ & $61.94 \pm 3.42$ & $58.17 \pm 3.41$ & $75.02 \pm 6.14$ & $73.44 \pm 5.82$ \\ 
 &  & \textbf{TS-TCC} & $77.83 \pm 6.90$ & $79.73 \pm 7.49$ & $76.18 \pm 7.21$ & $76.43 \pm 7.56$ & $84.12 \pm 7.32$ & $84.12 \pm 7.61$ \\ 
 &  & \textbf{SimCLR} & $56.09 \pm 2.25$ & $53.81 \pm 5.74$ & $51.73 \pm 2.39$ & $44.10 \pm 4.84$ & $53.81 \pm 5.74$ & $51.08 \pm 1.33$ \\  
 &  & \textbf{CLOCS} & $76.97 \pm 3.01$ & $81.70 \pm 3.21$ & $74.69 \pm 3.26$ & $74.75 \pm 3.61$ & $86.91 \pm 3.61$ & $86.70 \pm 3.64$ \\ 
 &  & \textbf{COMET} & $91.43 \pm 3.12$ & $92.52 \pm 2.36$ & $90.71 \pm 3.56$ & $91.14 \pm 3.31$ & $96.44 \pm 2.84$ & $96.48 \pm 2.82$ \\
&  & \textbf{COMET+DE} & $ \underline{93.47\pm 3.52}$ & $\underline{93.54 \pm 3.41}$ & $\underline{93.68 \pm 3.16}$ & $\underline{93.43 \pm 3.50}$ & $\textbf{98.27} \pm \textbf{1.34}$ & $\textbf{98.30} \pm \textbf{1.34}$ \\ 
&  & \textbf{COMET+ACL} & $ 92.77\pm 2.35$ & $92.92 \pm 2.43$ & $92.55 \pm 2.39$ & $92.66 \pm 2.37$ & $97.30 \pm 1.40$ & $97.39 \pm 1.40$ \\ 
&  & \textbf{DAAC} & $ \textbf{94.67}\pm \textbf{1.84}$ & $\textbf{94.93} \pm \textbf{1.76}$ & $\textbf{94.41} \pm \textbf{1.99}$ & $\textbf{94.58} \pm \textbf{1.89}$ & $\underline{98.10 \pm 1.20}$ & $\underline{98.19 \pm 1.15}$ \\ 
 \cline{2-9} \hline

\multirow{20}{*}{\textbf{TDBrain}} & \multirow{10}{*}{\textbf{100\%}} & \textbf{TS2vec} & $80.21 \pm 1.69$ & $81.38 \pm 1.97$ & $80.21 \pm 1.69$ & $80.07 \pm 1.69$ & $89.57 \pm 2.31$ & $89.60 \pm 2.37$ \\
 &  & \textbf{TF-C} & $66.62 \pm 1.76$ & $67.15 \pm 1.64$ & $66.62 \pm 1.76$ & $66.35 \pm 1.91$ & $65.43 \pm 6.13$ & $66.18 \pm 4.90$ \\ 
 &  & \textbf{Mixing-up} & $81.47 \pm 1.07$ & $82.11 \pm 1.12$ & $81.47 \pm 1.07$ & $81.38 \pm 1.08$ & $90.48 \pm 0.89$ & $90.51 \pm 0.04$ \\ 
 &  & \textbf{TS-TCC} & $77.42 \pm 2.86$ & $77.68 \pm 2.93$ & $77.42 \pm 2.86$ & $77.37 \pm 2.86$ & $87.37 \pm 3.06$ & $87.61 \pm 3.22$ \\ 
 &  & \textbf{SimCLR} & $58.43 \pm 1.77$ & $59.48 \pm 1.95$ & $58.43 \pm 1.77$ & $57.30 \pm 2.07$ & $59.48 \pm 1.95$ & $55.05 \pm 1.18$ \\ 
 &  & \textbf{CLOCS} & $78.16 \pm 1.13$ & $79.49 \pm 1.61$ & $78.16 \pm 1.13$ & $77.91 \pm 1.12$ & $88.49 \pm 1.95$ & $88.83 \pm 1.94$ \\ 
 &  & \textbf{COMET} & $85.47 \pm 1.16$ & $85.68 \pm 1.20$ & $85.47 \pm 1.16$ & $85.45 \pm 1.16$ & $93.73 \pm 1.02$ & $93.96 \pm 0.99$ \\ 
   &  & \textbf{COMET+DE} & $ 90.61\pm 1.7$ & $90.56 \pm 1.49$ & $90.16 \pm 1.94$ & $91.55 \pm 2.38$ & $95.69\pm 2.07$ & $96.07 \pm 1.85$ \\ 
 &  & \textbf{COMET+ACL} & $ \underline{93.26\pm 2.33}$ & $\underline{92.84 \pm 1.76}$ & $\underline{93.44 \pm 1.62}$ & $\underline{93.07 \pm 1.92}$ & $\underline{96.74 \pm 1.98}$ & $\underline{97.60 \pm 1.82}$ \\

 &  & \textbf{DAAC} & $ \textbf{95.60}\pm \textbf{1.20}$ & $\textbf{94.95} \pm \textbf{1.69}$ & $\textbf{95.73} \pm \textbf{1.74}$ & $\textbf{94.61} \pm \textbf{1.26}$ & $\textbf{97.63} \pm \textbf{1.57}$ & $\textbf{98.51} \pm \textbf{1.29}$ \\ \cline{2-9} 
 & \multirow{10}{*}{\textbf{10\%}} & \textbf{TS2vec} & $72.39 \pm 1.13$ & $74.49 \pm 1.73$ & $72.39 \pm 1.13$ & $71.80 \pm 1.05$ & $80.71 \pm 1.90$ & $80.06 \pm 1.87$ \\ 
 &  & \textbf{TF-C} & $59.14 \pm 3.04$ & $59.34 \pm 3.19$ & $59.14 \pm 3.04$ & $58.98 \pm 2.94$ & $59.56 \pm 4.10$ & $59.65 \pm 2.99$ \\
 &  & \textbf{Mixing-up} & $77.50 \pm 2.07$ & $80.09 \pm 1.92$ & $77.50 \pm 2.07$ & $76.99 \pm 2.28$ & $87.29 \pm 1.34$ & $87.13 \pm 1.37$ \\ 
 &  & \textbf{TS-TCC} & $71.23 \pm 1.57$ & $78.78 \pm 0.66$ & $71.23 \pm 1.57$ & $69.18 \pm 2.25$ & $80.56 \pm 1.98$ & $80.21 \pm 2.21$ \\ 
 &  & \textbf{SimCLR} & $59.79 \pm 2.09$ & $60.50 \pm 1.90$ & $59.79 \pm 2.09$ & $59.06 \pm 2.82$ & $60.50 \pm 1.90$ & $55.96 \pm 1.41$ \\  
 &  & \textbf{CLOCS} & $75.04 \pm 2.65$ & $75.97 \pm 2.86$ & $75.04 \pm 2.65$ & $74.83 \pm 2.66$ & $84.25 \pm 3.27$ & $84.37 \pm 3.52$ \\ 
 &  & \textbf{COMET} & $79.28 \pm 4.64$ & $79.83 \pm 4.83$ & $79.28 \pm 4.64$ & $79.19 \pm 4.62$ & $88.39 \pm 4.13$ & $88.38 \pm 3.96$ \\
 &  & \textbf{COMET+DE} & $ 81.27\pm 4.32$ & $\underline{81.92 \pm 4.29}$ & $\underline{82.30 \pm 4.33}$ & $82.07 \pm 4.52$ & $90.22 \pm 4.02$ & $90.50 \pm 4.09$ \\ 
  &  & \textbf{COMET+ACL} & $ \underline{82.20\pm3.08}$ & $81.53 \pm 3.91$ & $81.90 \pm 3.51$ & $\underline{82.09 \pm 3.60}$ & $\underline{93.73 \pm 3.27}$ & $\textbf{94.81} \pm \textbf{3.49}$ \\
 &  & \textbf{DAAC} & $ \textbf{83.66}\pm \textbf{3.29}$ & $\textbf{84.06} \pm \textbf{3.06}$ & $\textbf{83.29} \pm \textbf{2.96}$ & $\textbf{84.60} \pm \textbf{3.29}$ & $\textbf{95.27} \pm \textbf{4.01}$ & $\underline{94.71 \pm 4.09}$ \\ \cline{2-9} \hline
 
\multirow{20}{*}{\textbf{PTB}} & \multirow{10}{*}{\textbf{100\%}} & \textbf{TS2vec} & $85.14 \pm 1.66$ & $87.82 \pm 2.21$ & $76.84 \pm 3.99$ & $79.66 \pm 3.63$ & $90.50 \pm 1.59$ & $90.07 \pm 1.73$ \\
 &  & \textbf{TF-C} & $87.50 \pm 2.43$ & $85.50 \pm 3.04$ & $82.68 \pm 4.04$ & $83.77 \pm 3.50$ & $77.59 \pm 19.22$ & $80.62 \pm 15.10$ \\ 
 &  & \textbf{Mixing-up} & $87.61 \pm 1.48$ & $89.56 \pm 2.80$ & $79.30 \pm 1.67$ & $82.56 \pm 2.00$ & $89.29 \pm 1.26$ & $88.94 \pm 1.04$ \\ 
 &  & \textbf{TS-TCC} & $82.24 \pm 3.55$ & $85.28 \pm 5.08$ & $69.46 \pm 5.85$ & $72.08 \pm 6.85$ & $87.60 \pm 2.51$ & $86.26 \pm 3.00$ \\ 
 &  & \textbf{SimCLR} & $84.19 \pm 1.32$ & $85.85 \pm 1.89$ & $73.89 \pm 2.95$ & $76.84 \pm 2.80$ & $85.85 \pm 1.89$ & $70.70 \pm 2.36$ \\ 
 &  & \textbf{CLOCS} & $86.39 \pm 2.76$ & $87.06 \pm 2.81$ & $77.95 \pm 4.79$ & $80.71 \pm 4.78$ & $90.41 \pm 2.07$ & $87.35 \pm 3.36$ \\ 
 &  & \textbf{COMET} & $87.84 \pm 1.98$ & $87.67 \pm 1.72$ & $81.14 \pm 3.68$ & $83.45 \pm 3.22$ & $92.95 \pm 1.56$ & $87.47 \pm 2.82$ \\
  &  & \textbf{COMET+DE} & $88.49\pm2.06$ & $88.91 \pm 2.37$ & $83.49 \pm 4.59$ & $85.67 \pm 4.07$ & $93.88 \pm 1.94$ & $88.61 \pm 3.20$ \\ 
   &  & \textbf{COMET+ACL} & $ \underline{91.33\pm 2.03}$ & $\underline{90.89\pm2.83}$ & $\underline{84.20 \pm4.09}$ & $\underline{85.73\pm 3.82}$ & $\underline{94.16 \pm 2.27}$ & $\underline{90.02 \pm 3.05}$ \\
 &  & \textbf{DAAC} & $ \textbf{93.65}\pm \textbf{2.35}$ & $\textbf{93.28} \pm \textbf{1.86}$ & $\textbf{85.39}\pm \textbf{3.84}$ & $\textbf{87.64} \pm \textbf{3.27}$ & $\textbf{95.56} \pm \textbf{1.57}$ & $\textbf{90.28} \pm \textbf{3.49}$ \\ \cline{2-9} 
 & \multirow{10}{*}{\textbf{10\%}} & \textbf{TS2vec} & $82.49 \pm 4.71$ & $80.39 \pm 5.04$ & $83.35 \pm 0.91$ & $80.18 \pm 4.04$ & $93.03 \pm 1.03$ & $92.81 \pm 0.97$ \\
 &  & \textbf{TF-C} & $85.37 \pm 1.23$ & $82.80 \pm 2.35$ & $79.94 \pm 0.71$ & $81.09 \pm 1.14$ & $81.57 \pm 15.60$ & $84.57 \pm 8.12$ \\ 
 &  & \textbf{Mixing-up} & $87.05 \pm 1.41$ & $86.56 \pm 3.24$ & $80.61 \pm 2.68$ & $82.62 \pm 1.99$ & $89.28 \pm 1.43$ & $87.22 \pm 2.76$ \\ 
 &  & \textbf{TS-TCC} & $83.38 \pm 1.53$ & $85.11 \pm 2.11$ & $72.24 \pm 2.45$ & $75.27 \pm 2.64$ & $86.06 \pm 1.76$ & $84.34 \pm 2.08$ \\ 
 &  & \textbf{SimCLR} & $83.84 \pm 2.15$ & $87.19 \pm 1.34$ & $72.51 \pm 4.63$ & $75.44 \pm 4.77$ & $87.19 \pm 1.34$ & $69.99 \pm 3.84$ \\ 
 &  & \textbf{CLOCS} & $88.25 \pm 2.48$ & $88.64 \pm 2.12$ & $81.40 \pm 4.64$ & $83.84 \pm 4.03$ & $91.91 \pm 2.40$ & $89.76 \pm 3.94$ \\ 
 &  & \textbf{COMET} & $88.49 \pm 3.28$ & $88.98 \pm 2.60$ & $81.65 \pm 6.00$ & $84.01 \pm 5.61$ & $94.83 \pm 1.08$ & $92.48 \pm 2.22$ \\
 &  & \textbf{COMET+DE} & $ \underline{90.84\pm 3.01}$ & $\underline{89.67 \pm 2.88}$ & $83.05 \pm 2.89$ & $84.68 \pm 2.95$ & $95.20 \pm 2.62$ & $94.32 \pm 3.89$ \\ 
  &  & \textbf{COMET+ACL} & $ 89.67\pm 2.93$ & $89.66 \pm 2.47$ & $\underline{83.36 \pm 3.62}$ & $\underline{85.35 \pm 3.81}$ & $\underline{95.35 \pm 3.67}$ & $\underline{94.94 \pm 3.71}$ \\ 
& & \textbf{DAAC} & $ \textbf{91.35}\pm \textbf{2.77}$ & $\textbf{91.56} \pm \textbf{3.12}$ & $\textbf{85.23} \pm \textbf{2.65}$ & $\textbf{85.61} \pm \textbf{3.56}$ & $\textbf{96.30} \pm \textbf{3.22}$ & $\textbf{97.87} \pm \textbf{4.02}$ \\ \cline{2-9} \hline
\end{tabular}
}
\end{table}


\section{Conclusion}
In this work, we tackle two fundamental challenges in medical time-series diagnosis: the scarcity of labeled data and the limitations of conventional contrastive learning that rely on handcrafted sample pairs. We propose DAAC, a novel framework that enhances representation learning through a combination of external knowledge and adaptive contrastive strategies. By introducing a Discrepancy Estimator trained on large-scale normal data, we effectively quantify and incorporate abnormality as an auxiliary feature, reducing overfitting caused by limited target data. Additionally, our Adaptive Contrastive Learner, equipped with multi-head attention and hierarchical contrastive losses, enables the model to automatically discover meaningful relationships across diverse views and patient conditions without manual design. Together, these components allow DAAC to extract robust, generalizable features from complex medical time-series data. Experimental results on three diagnostic tasks confirm that DAAC achieves state-of-the-art performance, even with minimal labeled data, highlighting its potential for real-world deployment in low-resource clinical settings.

While DAAC demonstrates its effectiveness, the current discrepancy estimation may still be suboptimal in fully leveraging external information. Although care is taken to prevent external data from overwhelming the target dataset, the representation of discrepancy remains somewhat coarse. Future work could explore more precise and adaptive strategies for modeling discrepancy, aiming to better balance external knowledge integration with the inherent distribution of the target data. Additionally, we plan to extend DAAC to other modalities that share similar hierarchical or temporal structures, broadening its applicability. We also acknowledge that jointly optimizing all five loss weights may further improve performance, and we have included this direction as a key part of our future work.

\bibliographystyle{unsrt}  
\bibliography{references}  

\appendix

\section{Preliminary Knowledge on Medical Time Series}
\subsection{Hierarchical Structuring of Medical Time Series}
\label{sec:appendix_a_1}
To better understand the structure of medical time series data, this section provides a comprehensive explanation of its hierarchical organization, using EEG signals from Alzheimer’s disease as an example. The dataset is categorized into four hierarchical levels: Subject, Trial, Epoch, and Temporal, with each level representing a unique granularity of the data (Appendix figure~\ref{fig:sample_image}). Unlike conventional time series, which typically consist of sample and observation levels, medical time series incorporate two additional layers: patient and trial. These extra layers enhance the dataset's structure, enabling the development of methods tailored to the specific characteristics and analytical requirements of medical time series. As illustrated in Appendix Table~\ref{tab:models_levels}, many existing approaches utilize only a subset of these levels, underscoring the importance of leveraging the complete hierarchy for more effective analysis.

\textbf{Definition 1: Subject}

A \textbf{subject} \( \mathbf{p}_j \in \mathbb{R}^G \) in medical time series data represents the collection of multiple trials corresponding to a single individual. The subscript \( j \) denotes the subject ID. Trials associated with the same subject may differ due to variations in data collection timeframes, sensor placement, or patient conditions. A subject's data is generally represented as an aggregate of trials. Each trial within a subject \( \mathbf{p}_j \) is typically segmented into smaller samples to facilitate representation learning. To denote all the samples belonging to a specific subject \( \mathbf{p}_j \), we use the notation \( \mathbf{P}_j \).

\textbf{Definition 2: Trial}

A \textbf{trial} \( \mathbf{r}_k \) refers to a continuous set of observations collected over an extended time period (e.g., 30 minutes). This can also be referred to as a \textbf{record}. The subscript \( k \) denotes the trial ID. Trials are generally too lengthy (e.g., consisting of hundreds of thousands of observations) to be directly processed by deep learning models. Therefore, they are usually divided into smaller subsequences, referred to as samples or segments. The collection of all samples derived from a trial \( \mathbf{r}_k \) is denoted by \( \mathbf{R}_k \).

\textbf{Definition 3: Epoch}

An \textbf{epoch} \( e_m \) is a sequence of consecutive observations derived from a trial, typically spanning a shorter time period (e.g., several seconds). In EEG (electroencephalography), an epoch refers to a specific time window extracted from the continuous EEG signal. These time windows are used to isolate signal segments for further analysis, often aligned with specific events or experimental conditions. Each epoch is composed of multiple observations measured at regular intervals over a time range defined by \( M \) timestamps. To denote a specific epoch within a trial, we use \( e_m \), where \( m \) represents the epoch index. The set of observations forming an epoch is \[e_m = \{ e_{m,t} \mid t = 1, \cdots, M \}.\]

\textbf{Definition 4: Temporal}

A \textbf{temporal observation} \( \mathbf{x}_{n,t} \in \mathbb{R}^H \) refers to a single data point or vector recorded at a specific timestamp \( t \). The subscript \( n \) denotes the sample index, while \( t \) represents the timestamp. For univariate time series, the temporal observation is a single real value, whereas, for multivariate time series, it is a vector with \( H \) dimensions. Temporal observations may represent physiological signals, laboratory measurements, or other health-related indicators.

\subsection{Data Scarcity and Heterogeneity for Medical Time Series}
\label{sec:appendix_a_2}

\textbf{Data Scarcity}
Medical data, especially high-quality neurodegenerative disease data, is often expensive, with high costs associated with data collection and annotation. Additionally, individual datasets typically have limited sample sizes and suffer from class imbalance issues (e.g., an unequal ratio of healthy individuals to patients). These problems lead to overfitting during model training, poor generalization, and suboptimal classification performance.

\textbf{Data Heterogeneity}
Medical data often exhibits a multi-center characteristic, meaning datasets from different institutions show significant distributional differences. This heterogeneity makes it difficult to directly transfer models trained on one dataset to others, severely hindering the development and application of robust deep learning models.

\section{Supplementary Figures and Tables}
\label{sec:appendix_b}
\renewcommand{\thetable}{B\arabic{table}}
\renewcommand{\thefigure}{B\arabic{figure}} 
\setcounter{table}{0}
\setcounter{figure}{0}



\begin{table}[htbp]
\centering
\caption{\textbf{Different Models Utilize Different Levels}}
\label{tab:models_levels}
\small
\setlength{\tabcolsep}{6pt} 
\begin{tabular}{lccccc}
\hline
\textbf{Models} & \textbf{Subject} & \textbf{Trial} & \textbf{Epoch} & \textbf{Temporal} & \textbf{View} \\
\hline
\textbf{SimCLR} &           &           & \checkmark &           &           \\
\textbf{TF-C}   &           &           & \checkmark &           &           \\
\textbf{Mixing-up} &        &           & \checkmark &           &           \\
\textbf{TNC}    &           & \checkmark &           &           &           \\
\textbf{TS2vec} &           &           & \checkmark & \checkmark &      \\
\textbf{TS-TCC} &           &           & \checkmark & \checkmark &      \\
\textbf{CLOCS}  & \checkmark &          & \checkmark &           &      \\
\textbf{COMET}  & \checkmark & \checkmark & \checkmark & \checkmark & \\
\hline
\textbf{DAAC}   & \checkmark & \checkmark & \checkmark & \checkmark & \textbf{\checkmark} \\
\hline
\end{tabular}
\end{table}

\begin{figure}[htbp]
    \centering

    \includegraphics[width=0.8\textwidth]{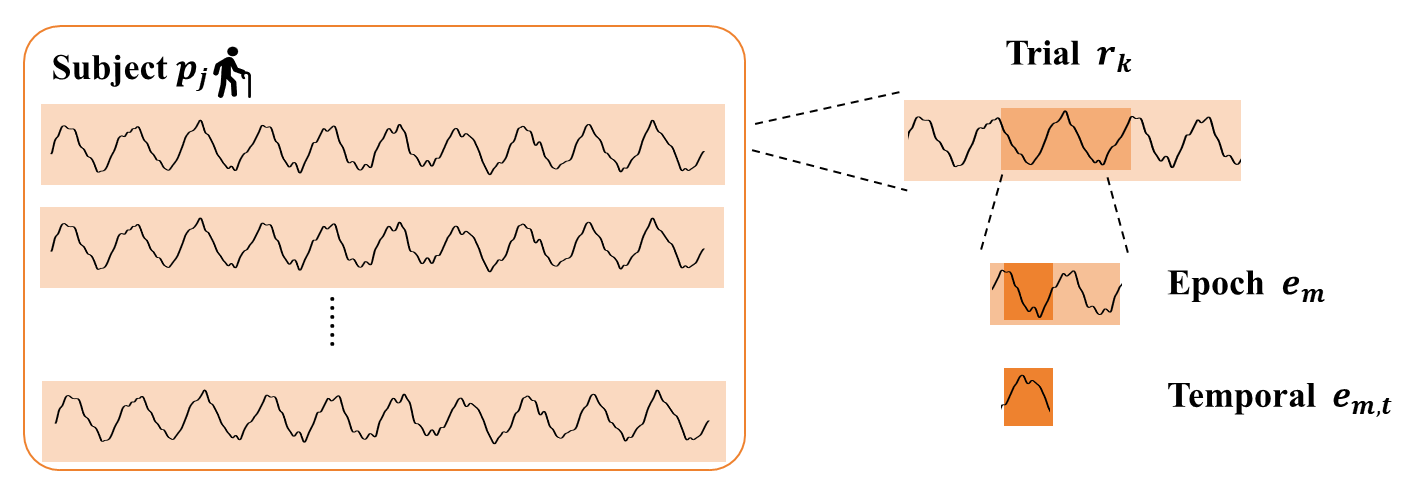} 
    \caption{\textbf{Structure of Medical Time Series.}The data is organized into four levels: Subject (\( p_j \)), Trial (\( r_k \)), Epoch (\( e_m \)), and Temporal (\( e_{m,t} \)), capturing different granularities of the data, from subject-level recordings to temporal observations within individual epochs.}
    \label{fig:sample_image}
\end{figure}

\begin{figure}[htbp]
    \centering
    \includegraphics[width=0.8\textwidth]{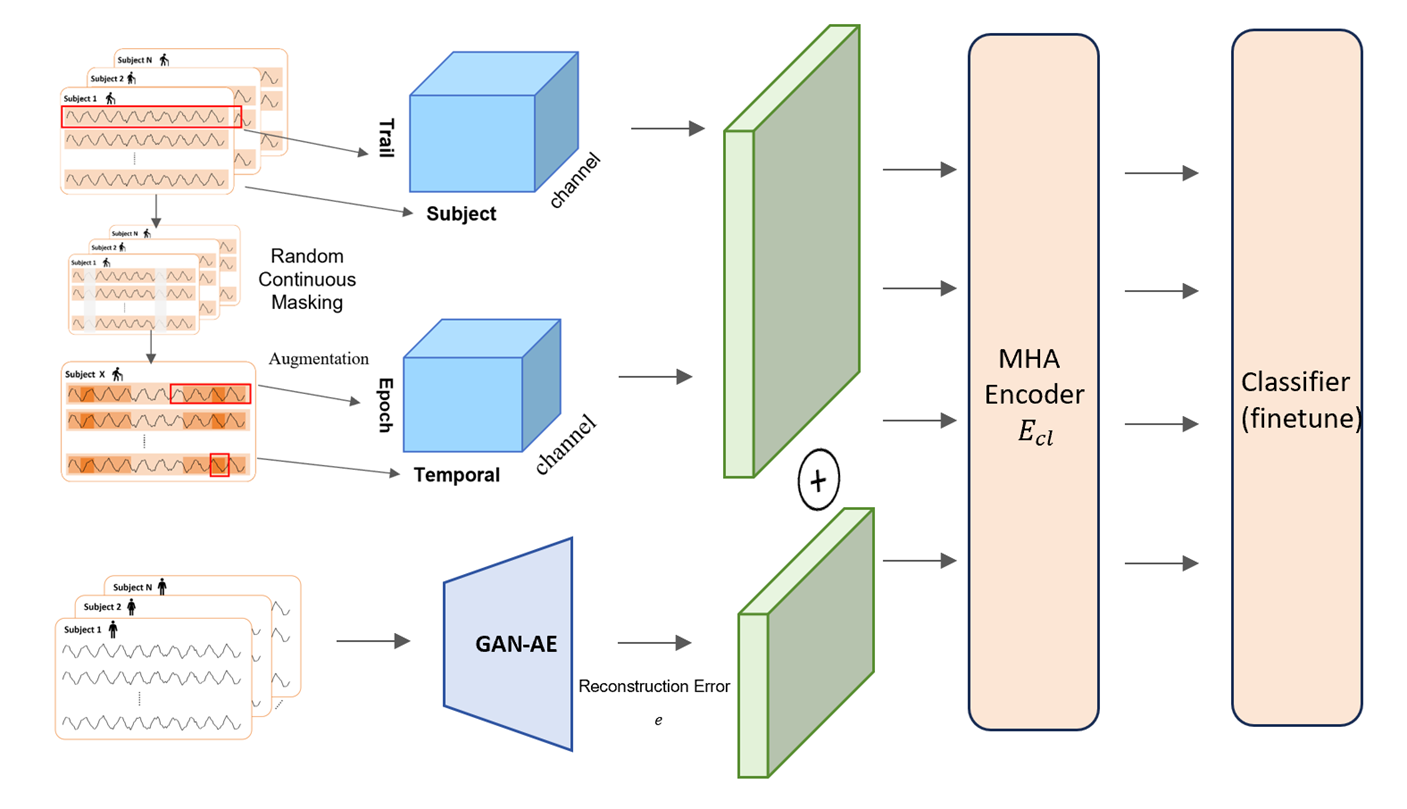} 
    \caption{Overview of DAAC}
    \label{fig:overview}
\end{figure}


\section{Datasets and Data Preprocessing}
\renewcommand{\thetable}{C\arabic{table}}
\renewcommand{\thefigure}{C\arabic{figure}} 
\setcounter{table}{0}
\setcounter{figure}{0}

\label{sec:appendix_c}
In the medical field, datasets from different hospitals often exhibit a certain degree of distributional shift. Models trained on data from one hospital may demonstrate unstable performance when applied to others. To enhance model generalization, we propose leveraging DE to incorporate data from other hospitals. Although distributional differences exist, we argue that the sequential patterns of normal data (i.e., data from healthy patients) can still provide valuable guidance for diagnosis in the target hospital.

In this study:

The dataset from the current hospital is referred to as the \textbf{internal dataset (or target dataset)}, which participates in the second and third stages of model training.
Datasets from other hospitals, used for guidance, are termed \textbf{external datasets which are from healthy patients} and contribute to the first stage of training.

\renewcommand{\thesection}{C\arabic{section}} 
\renewcommand{\thesubsection}
{C\arabic{section}.\arabic{subsection}} 
\setcounter{section}{0} 

\section{External Datasets} 
\subsection{Alzheimer’s Disease Dataset}
The AD dataset~\cite{miltiadous2023dataset} is a publicly available EEG time-series dataset comprising three classes: 36 Alzheimer's disease (AD) patients, 23 frontotemporal dementia (FTD) patients, and 29 healthy control (HC) subjects. Recordings were obtained using 19 channels at a raw sampling rate of 500~Hz, with each subject contributing a single trial. The average trial durations were approximately 13.5 minutes for AD subjects (range: 5.1--21.3 minutes), 12 minutes for FTD subjects (range: 7.9--16.9 minutes), and 13.8 minutes for HC subjects (range: 12.5--16.5 minutes). A bandpass filter between 0.5--45~Hz was applied to all recordings. Subsequently, the data were downsampled to 256~Hz and segmented into non-overlapping 1-second windows (256 timestamps per segment), discarding any segments shorter than 1 second. Sixteen overlapping channels were selected from the original 19 channels for analysis. After preprocessing, the dataset contained a total of 69,752 samples. We considered both subject-dependent and subject-independent evaluation setups: in the subject-dependent setup, 60\%, 20\%, and 20\% of the samples were allocated to the training, validation, and test sets, respectively; in the subject-independent setup, 60\%, 20\%, and 20\% of the subjects (and their corresponding samples) were used for training, validation, and testing, respectively.
\subsection{Parkinson's Disease Dataset} 
The PD dataset used in this study consists of EEG recordings from 46 participants collected at the University of New Mexico (UNM, Albuquerque, New Mexico), as previously described in~\cite{anjum2020linear}. The cohort includes 22 individuals diagnosed with Parkinson’s disease (PD) in an OFF-medication state and 24 healthy control (HC) subjects. EEG recordings were obtained using a 64-channel Brain Vision system at a sampling rate of 500~Hz. For consistency across datasets, 33 channels with matched naming conventions were selected for analysis. The EEG signals were high-pass filtered at 1~Hz to remove low-frequency noise, and no further preprocessing was applied. Only the first one minute of eyes-closed resting-state EEG data was used for each participant. Each 1-minute recording was segmented into non-overlapping 5-second windows, resulting in data represented as \(\mathbb{R}^{n \times 33 \times 1250}\), where \(n\) denotes the number of 5-second segments and 1250 is the number of timestamps per segment.
\subsection{Myocardial Infarction Dataset} 
The PTB-XL dataset~\cite{wagner2020ptb} is a large-scale public ECG time-series dataset comprising recordings from 18,869 subjects across 12 channels, with five labels corresponding to four heart disease categories and one healthy control (HC) category. Each subject may contribute one or more trials; however, to ensure consistency, subjects with varying diagnostic results across trials were excluded, resulting in a final dataset of 17,596 subjects. Recordings were obtained as 10-second intervals and are available in two versions with sampling frequencies of 100~Hz and 500~Hz. In this study, we use the 500~Hz version, which is downsampled to 250~Hz and normalized using standard scalers. Each trial was then segmented into non-overlapping 1-second windows (250 timestamps per segment), discarding any segments shorter than 1 second. After preprocessing, the dataset contained 191,400 samples. A subject-independent setup was adopted for dataset splitting, allocating 60\%, 20\%, and 20\% of the subjects (and their corresponding samples) to the training, validation, and test sets, respectively.
\section{Target Datasets}
\subsection{Alzheimer’s Disease Dataset}
The AD dataset~\cite{escudero2006analysis} consists of EEG recordings from 22 subjects, including 11 patients diagnosed with probable Alzheimer’s disease (5 men, 6 women; mean age \(72.5 \pm 8.3\) years) and 11 healthy controls. Patients were recruited from the Alzheimer’s Patients’ Relatives Association of Valladolid (AFAVA) and underwent comprehensive clinical evaluations, including neurological and physical examinations, brain imaging, and Mini-Mental State Examination (MMSE) assessments. EEG recordings were conducted at the University Hospital of Valladolid (Spain) using Profile Study Room 2.3.411 EEG equipment (Oxford Instruments) with 19 electrode positions according to the international 10--20 system. Each subject contributed over 5 minutes of EEG data recorded in a relaxed, eyes-closed, awake state to minimize artifacts. Signals were sampled at 256~Hz with 12-bit analog-to-digital precision.Each patient contributed an average of \(30.0 \pm 12.5\) trials, with each trial spanning a 5-second interval across 16 channels. Prior to analysis, trials were standardized using a standard scaler and segmented into nine half-overlapping 1-second samples, each containing 256 timestamps.

\subsection{Parkinson's Disease Dataset} 
The TDBrain dataset~\cite{van2022two} contains EEG recordings from 1,274 patients acquired across 33 channels at 500~Hz during eyes-closed (EC) and eyes-open (EO) tasks, covering 60 distinct neurological conditions, with some patients diagnosed with multiple disorders. In this study, we focus on a subset comprising 25 Parkinson’s disease (PD) patients and 25 healthy controls, utilizing only the EC task for representation learning.
Raw EC trials were preprocessed using a sliding window approach, where each trial was resampled to 256~Hz and divided into pseudo-trials of 2,560 timestamps (10 seconds). Each pseudo-trial was standardized and subsequently segmented into 19 half-overlapping 1-second windows, each containing 256 timestamps. Each sample was assigned a binary label (PD or healthy), along with patient and trial identifiers, where trial identifiers correspond to the generated pseudo-trials.A patient-independent split was employed, with samples from eight patients (four PD and four healthy) allocated to the validation set, another eight patients to the test set, and the remaining subjects to the training set.

\subsection{Myocardial Infarction Dataset}
The PTB dataset~\cite{bousseljot1995nutzung} contains ECG recordings from 290 patients across 15 channels, sampled at 1,000~Hz, and covers eight types of heart diseases. In this study, we focus on a subset comprising 198 patients for a binary classification task, specifically distinguishing between myocardial infarction (MI) and healthy controls. The ECG signals were resampled to 250~Hz, normalized using a standard scaler, and segmented into individual heartbeats to preserve critical peak information, as a standard sliding window approach may lead to information loss.The segmentation process involved several steps: (1) determining the maximum heartbeat duration by calculating the median R-peak interval across all channels in each trial, excluding outliers; (2) identifying the position of the first R-peak using the median value across channels; (3) splitting raw trials into individual heartbeats based on the median R-peak interval, with segments extended symmetrically around the R-peak; and (4) applying zero-padding to ensure uniform sample lengths.A patient-independent split was adopted, with samples from 28 patients (7 healthy and 21 MI) allocated to the validation and test sets, and the remaining subjects used for training.

\section{Domain Shift Study}
Although we hope that the external normal data assumes a certain level of domain alignment. In practice, however, our external and target datasets exhibit notable differences, including variations in population demographics and acquisition protocols.

For example, in the AD task, the external dataset (~\cite{miltiadous2023dataset}) has an average subject age of 63.6 years, while the target dataset (~\cite{escudero2006analysis}) has an average age of 72.5 years. When training the Discrepancy Estimator (DE) using this external dataset, we indeed observe signs of negative transfer—specifically, in Table 2, the AUROC of COMET+DE drops from 94.44 to 93.97.

Despite this, DAAC still outperforms both COMET+ACL and COMET+DE in the same setting, as shown in Table 2. This suggests that the Adaptive Contrastive Learner (ACL) effectively mitigates the potential negative impact introduced by domain-shifted external data, and in some cases, even corrects for suboptimal guidance from the DE. This robustness highlights the strength of DAAC’s multi-stage design in handling real-world domain variability.

\renewcommand{\thesection}{D}

\renewcommand{\thesubsection}{D.\arabic{subsection}}
\renewcommand{\thetable}{D\arabic{table}}
\renewcommand{\thefigure}{D\arabic{figure}} 
\setcounter{table}{0}
\setcounter{figure}{0}

\setcounter{section}{0}
\setcounter{subsection}{0}
\section{Loss Setups}
\label{sec:appendix_d}
\subsection{Loss Funtion of DE}
The model is trained by optimizing two loss functions: the discriminator loss and the generator loss. During the training process, the discriminator is first updated using the discriminator loss ($L_D$), followed by the generator. The two components are updated alternately.

\textbf{(1) Discriminator Loss}:
  \[
  L_D = \frac{1}{2} (L_{D_{\text{real}}} + L_{D_{\text{fake}}})
  \]
  where $L_{D_{\text{real}}} = \text{BCE}(C(xx_i), 1)$ indicates the loss for real data, encouraging the discriminator to classify real data as 1. $L_{D_{\text{fake}}} = \text{BCE}(C(\hat{xx}_i), 0)$ represents the loss for generated data, encouraging the discriminator to classify generated data as 0. $\text{BCE}$ is the binary cross-entropy loss function.

\textbf{(2) Generator Loss}:
  \[
  L_G = L_{G_{\text{adv}}} + L_{G_{\text{recon}}}
  \]
  where $L_{G_{\text{adv}}} = \text{BCE}(C(\hat{xx}_i), 1)$ indicates The adversarial loss, encouraging the discriminator to classify generated data as 1. $L_{G_{\text{recon}}} = \text{MSE}(\hat{xx}_i, xx_i)$ represents the reconstruction loss, ensuring that the generated data $\hat{xx}_i$ is close to the original data $xx_i$. $\text{MSE}$ is the mean squared error loss function.
  
\subsection{Loss Weight Sensitivity}
\label{Loss Weight Sensitivity}

\section*{Appendix G.2 Loss Weight Sensitivity Analysis}

To evaluate the robustness and effectiveness of our loss function design, we performed a comprehensive sensitivity analysis on the weighting strategy of its five components. The total loss is a weighted sum of four hierarchical contrastive losses and one view contrastive loss (\(L_V\)). In our main experiments, we used a fixed configuration of \(1{:}1{:}1{:}1{:}2\), which we found to yield consistent robust performance across datasets.

The first four hierarchical components were assigned equal weights (1:1:1:1), following prior practices (~\cite{wang2024contrast}), which we hypothesized to provide a strong default setting. To test this assumption, we individually perturbed each of these four weights while keeping the others fixed, and evaluated the model on three datasets (AD, PTB, TDBrain) under both 100\% and 10\% data regimes. Additionally, to investigate the contribution of the proposed ACL, we varied the weight of \(L_V\) from 0 to 3 and monitored the impact on overall performance.

The full results are provided in Table~\ref{tab:loss weight}. We summarize the main findings as follows:

(1) \textbf{Effectiveness of \(L_V\):} Removing \(L_V\) (i.e., setting its weight to 0) consistently led to substantial performance degradation across all evaluation metrics and datasets, highlighting its critical role in enhancing cross-view representation alignment.

(2) \textbf{Performance saturation with increasing \(L_V\):} Increasing the weight of \(L_V\) from 1 to 2 led to noticeable gains in accuracy, F1 score, AUROC, and AUPRC. However, further increasing the weight to 3 resulted in only marginal or dataset-specific improvements, suggesting a saturation effect.

(3) \textbf{Default weight robustness:} Perturbing the weights of the first four components (e.g., 1:1:2:1:1, 1:2:1:1:1) produced only minor variations in performance, indicating that the default configuration (1:1:1:1) is a generally robust and effective setting.

These findings provide empirical justification for our selected weight configuration and demonstrate that our method performs reliably without the need for extensive tuning. Further improvements may be achievable through joint or dynamic weight optimization, which we leave as future work.

\begin{table}
\centering
\scriptsize
\caption{Comparison of performance metrics under different hyperparameter configurations on AD, PTB, and TDBrain datasets.}
\label{tab:loss_weight}
\begin{tblr}{
  cells = {c},
  cell{2}{1} = {r=16}{},
  cell{2}{2} = {r=8}{},
  cell{18}{1} = {r=16}{},
  cell{18}{2} = {r=8}{},
  cell{34}{1} = {r=16}{},
  cell{34}{2} = {r=8}{},
  hline{1-2,18,34,50} = {-}{},
  hline{3-9,11-17,19-25,27-33,35-41,43-49} = {3-9}{},
  hline{10,26,42} = {2-9}{},
}
Dataset & Fraction & Config    & Accuracy                & Precision               & Recall                  & F1 score                & AUROC                   & AUPRC                   \\
AD      & 100      & 1,1,1,1,0 & 84.50$\pm$4.46          & 88.31$\pm$2.42          & 82.95$\pm$5.39          & 83.33$\pm$5.15          & 93.60$\pm$2.37          & 94.10$\pm$2.48          \\
        &          & 1,1,1,1,1 & 88.50$\pm$4.20          & 90.20$\pm$3.60          & 87.30$\pm$4.90          & 88.10$\pm$4.60          & 95.50$\pm$2.10          & 95.60$\pm$2.00          \\
        &          & 1,1,1,1,2 & \textbf{91.16$\pm$4.14} & \textbf{92.10$\pm$3.66} & \uline{90.50$\pm$4.51}  & \uline{90.88$\pm$4.33}  & \textbf{96.40$\pm$2.76} & \textbf{96.30$\pm$2.82} \\
        &          & 1,1,1,1,3 & \uline{91.02$\pm$3.15}  & \uline{92.06$\pm$3.54}  & \textbf{90.58$\pm$5.11} & \textbf{90.98$\pm$3.98} & \uline{96.24$\pm$2.88}  & \uline{95.63$\pm$2.43}  \\
        &          & 1,1,1,2,1 & 87.20$\pm$4.30          & 89.30$\pm$4.00          & 86.80$\pm$4.60          & 87.30$\pm$4.30          & 94.65$\pm$2.90          & 94.75$\pm$2.80          \\
        &          & 1,1,2,1,1 & 87.40$\pm$4.30          & 89.50$\pm$4.10          & 86.90$\pm$4.50          & 87.40$\pm$4.20          & 94.80$\pm$2.85          & 94.90$\pm$2.95          \\
        &          & 1,2,1,1,1 & 86.90$\pm$4.60          & 89.20$\pm$4.00          & 86.50$\pm$4.90          & 87.10$\pm$4.50          & 94.40$\pm$2.90          & 94.50$\pm$2.80          \\
        &          & 2,1,1,1,1 & 86.60$\pm$4.70          & 88.70$\pm$4.20          & 86.20$\pm$5.10          & 86.80$\pm$4.60          & 93.90$\pm$3.10          & 94.00$\pm$3.00          \\
        &          & 1,1,1,1,0 & 91.43$\pm$3.12          & 92.52$\pm$2.36          & 90.71$\pm$3.56          & 91.14$\pm$3.31          & 92.20$\pm$2.84          & 92.30$\pm$2.82          \\
        &          & 1,1,1,1,1 & 92.30$\pm$2.20          & \uline{92.80$\pm$2.10}  & 92.10$\pm$2.50          & \uline{92.20$\pm$2.45}  & 96.70$\pm$1.90          & 96.60$\pm$1.95          \\
        &          & 1,1,1,1,2 & \uline{92.77$\pm$2.35}  & \textbf{92.92$\pm$2.43} & \uline{92.55$\pm$2.39}  & \textbf{92.66$\pm$2.37} & \textbf{97.40$\pm$1.40} & \uline{97.50$\pm$1.45}  \\
        & 10       & 1,1,1,1,3 & \textbf{92.86$\pm$2.44} & 92.33$\pm$3.61          & \textbf{92.63$\pm$2.01} & 91.54$\pm$3.03          & \uline{96.88$\pm$1.89}  & \textbf{97.64$\pm$1.33} \\
        &          & 1,1,1,2,1 & 92.00$\pm$2.80          & 92.50$\pm$2.50          & 91.90$\pm$2.70          & 92.10$\pm$2.60          & 95.20$\pm$2.00          & 95.10$\pm$2.05          \\
        &          & 1,1,2,1,1 & 91.90$\pm$2.90          & 92.40$\pm$2.55          & 91.80$\pm$2.65          & 92.00$\pm$2.60          & 94.50$\pm$2.10          & 94.40$\pm$2.05          \\
        &          & 1,2,1,1,1 & 91.70$\pm$3.00          & 92.10$\pm$2.80          & 91.60$\pm$2.75          & 91.90$\pm$2.70          & 94.00$\pm$2.20          & 94.10$\pm$2.15          \\
        &          & 2,1,1,1,1 & 91.60$\pm$3.05          & 92.00$\pm$2.90          & 91.50$\pm$2.80          & 91.80$\pm$2.75          & 93.40$\pm$2.30          & 93.50$\pm$2.25          \\
PTB     & 100      & 1,1,1,1,0 & 86.36$\pm$1.44          & 87.18$\pm$1.28          & 77.87$\pm$2.55          & 80.79$\pm$2.44          & 92.60$\pm$2.35          & 89.09$\pm$1.64          \\
        &          & 1,1,1,1,1 & 89.70$\pm$1.80          & 89.60$\pm$2.10          & 82.80$\pm$3.00          & 85.80$\pm$2.90          & 94.00$\pm$2.10          & \uline{90.10$\pm$2.40}  \\
        &          & 1,1,1,1,2 & \uline{91.33$\pm$2.03}  & \textbf{90.89$\pm$2.83} & \textbf{84.20$\pm$4.09} & \textbf{87.64$\pm$3.82} & \textbf{94.16$\pm$2.27} & 90.02$\pm$3.05          \\
        &          & 1,1,1,1,3 & \textbf{91.46$\pm$1.97} & \uline{89.86$\pm$2.43}  & \uline{83.56$\pm$3.68}  & \uline{87.38$\pm$4.10}  & \uline{94.06$\pm$2.43}  & \textbf{90.11$\pm$2.71} \\
        &          & 1,1,1,2,1 & 89.00$\pm$2.00          & 89.30$\pm$2.20          & 82.10$\pm$2.90          & 85.30$\pm$3.00          & 93.70$\pm$2.20          & 89.90$\pm$2.50          \\
        &          & 1,1,2,1,1 & 88.60$\pm$2.10          & 88.90$\pm$2.30          & 81.70$\pm$3.10          & 84.80$\pm$3.10          & 93.45$\pm$2.15          & 89.70$\pm$2.40          \\
        &          & 1,2,1,1,1 & 88.10$\pm$2.20          & 88.30$\pm$2.40          & 81.20$\pm$3.30          & 84.10$\pm$3.20          & 93.20$\pm$2.25          & 89.60$\pm$2.35          \\
        &          & 2,1,1,1,1 & 87.80$\pm$2.30          & 87.90$\pm$2.50          & 80.70$\pm$3.40          & 83.80$\pm$3.30          & 92.90$\pm$2.30          & 89.40$\pm$2.30          \\
        &          & 1,1,1,1,0 & 90.32$\pm$1.61          & \uline{89.67$\pm$1.94}  & \textbf{85.85$\pm$2.99} & \textbf{87.35$\pm$2.36} & 91.30$\pm$1.11          & 90.46$\pm$3.09          \\
        &          & 1,1,1,1,1 & \textbf{90.90$\pm$2.00} & \textbf{89.90$\pm$2.20} & \uline{84.60$\pm$3.00}  & \uline{86.30$\pm$2.80}  & 94.10$\pm$2.30          & 92.20$\pm$3.10          \\
        &          & 1,1,1,1,2 & 89.67$\pm$2.93          & 89.66$\pm$2.47          & 83.36$\pm$3.62          & 85.35$\pm$3.81          & \uline{95.35$\pm$3.67}  & \uline{94.94$\pm$3.71}  \\
        &          & 1,1,1,1,3 & 89.77$\pm$2.62          & 88.97$\pm$3.01          & 84.04$\pm$3.42          & 86.15$\pm$4.22          & \textbf{95.77$\pm$4.11} & \textbf{95.01$\pm$3.50} \\
        & 10       & 1,1,1,2,1 & \uline{90.60$\pm$1.90}  & 89.80$\pm$2.10          & 84.20$\pm$3.10          & 86.00$\pm$2.90          & 93.40$\pm$2.50          & 91.90$\pm$3.00          \\
        &          & 1,1,2,1,1 & 90.50$\pm$2.10          & 89.60$\pm$2.00          & 84.00$\pm$3.20          & 85.80$\pm$3.00          & 92.90$\pm$2.60          & 91.70$\pm$2.80          \\
        &          & 1,2,1,1,1 & 90.10$\pm$2.20          & 89.50$\pm$2.20          & 83.80$\pm$3.40          & 85.60$\pm$3.20          & 92.50$\pm$2.80          & 91.60$\pm$2.70          \\
        &          & 2,1,1,1,1 & 89.70$\pm$2.30          & 89.30$\pm$2.30          & 83.50$\pm$3.50          & 85.10$\pm$3.40          & 91.80$\pm$3.00          & 91.10$\pm$2.80          \\
TDBrain & 100      & 1,1,1,1,0 & 85.47$\pm$1.16          & 85.68$\pm$1.20          & 85.47$\pm$1.16          & 85.45$\pm$1.16          & 91.80$\pm$1.02          & 93.96$\pm$0.99          \\
        &          & 1,1,1,1,1 & 91.20$\pm$1.90          & 90.60$\pm$1.80          & 90.90$\pm$1.75          & 91.10$\pm$1.85          & 95.60$\pm$1.60          & 96.20$\pm$1.65          \\
        &          & 1,1,1,1,2 & \uline{93.26$\pm$2.33}  & \textbf{92.84$\pm$1.76} & \textbf{93.44$\pm$1.62} & \uline{93.07$\pm$1.92}  & \textbf{96.74$\pm$1.98} & \textbf{97.60$\pm$1.82} \\
        &          & 1,1,1,1,3 & \textbf{93.46$\pm$2.83} & \uline{92.49$\pm$2.07}  & \uline{93.04$\pm$1.34}  & \textbf{93.43$\pm$2.13} & \uline{96.39$\pm$1.56}  & \uline{97.44$\pm$1.38}  \\
        &          & 1,1,1,2,1 & 90.70$\pm$1.85          & 90.30$\pm$1.95          & 90.60$\pm$1.80          & 90.80$\pm$1.90          & 94.80$\pm$1.75          & 95.60$\pm$1.70          \\
        &          & 1,1,2,1,1 & 90.40$\pm$1.95          & 90.00$\pm$2.05          & 90.20$\pm$1.85          & 90.50$\pm$1.95          & 94.10$\pm$1.80          & 95.20$\pm$1.75          \\
        &          & 1,2,1,1,1 & 89.90$\pm$2.05          & 89.70$\pm$2.15          & 89.80$\pm$2.00          & 90.00$\pm$2.10          & 93.30$\pm$1.85          & 94.50$\pm$1.80          \\
        &          & 2,1,1,1,1 & 89.40$\pm$2.10          & 89.20$\pm$2.25          & 89.50$\pm$2.05          & 89.80$\pm$2.20          & 92.40$\pm$1.90          & 94.00$\pm$1.85          \\
        &          & 1,1,1,1,0 & 79.28$\pm$4.64          & 79.83$\pm$4.83          & 79.28$\pm$4.64          & 79.19$\pm$4.62          & 88.39$\pm$4.13          & 88.38$\pm$3.96          \\
        &          & 1,1,1,1,1 & 81.00$\pm$3.60          & 80.80$\pm$3.90          & 80.90$\pm$3.70          & 81.00$\pm$3.65          & 92.60$\pm$3.50          & 93.80$\pm$3.60          \\
        &          & 1,1,1,1,2 & \textbf{82.20$\pm$3.08} & \textbf{81.53$\pm$3.91} & \textbf{81.90$\pm$3.51} & \textbf{82.09$\pm$3.60} & \textbf{93.73$\pm$3.27} & \textbf{94.81$\pm$3.49} \\
        & 10       & 1,1,1,1,3 & \uline{81.49$\pm$2.97}  & \uline{81.07$\pm$3.27}  & \uline{81.45$\pm$2.90}  & \uline{81.27$\pm$3.73}  & \uline{93.40$\pm$2.95}  & \uline{94.01$\pm$2.94}  \\
        &          & 1,1,1,2,1 & 80.70$\pm$3.80          & 80.50$\pm$3.60          & 80.80$\pm$3.70          & 80.90$\pm$3.80          & 91.60$\pm$3.60          & 92.90$\pm$3.55          \\
        &          & 1,1,2,1,1 & 80.40$\pm$3.90          & 80.20$\pm$3.80          & 80.50$\pm$3.80          & 80.60$\pm$3.90          & 91.00$\pm$3.70          & 92.40$\pm$3.60          \\
        &          & 1,2,1,1,1 & 80.10$\pm$4.00          & 80.00$\pm$3.90          & 80.20$\pm$3.90          & 80.30$\pm$3.90          & 90.40$\pm$3.80          & 91.80$\pm$3.70          \\
        &          & 2,1,1,1,1 & 79.80$\pm$4.20          & 79.90$\pm$4.00          & 79.70$\pm$4.00          & 79.80$\pm$4.10          & 89.60$\pm$3.90          & 91.20$\pm$3.80          
\end{tblr}
\end{table}


\renewcommand{\thesection}{E}

\renewcommand{\thesubsection}{E.\arabic{subsection}}

\setcounter{section}{0}
\setcounter{subsection}{0}
\renewcommand{\thetable}{E\arabic{table}}
\renewcommand{\thefigure}{E\arabic{figure}} 
\setcounter{table}{0}
\setcounter{figure}{0}

\section{Ablation Study}
\label{sec:appendix_e}
\subsection{Mutual Information Analysis of Discrepancy Estimator Reconstruction Error}
We evaluate the statistical relevance of the Discrepancy Estimator (DE) derived reconstruction error by computing mutual information (MI) scores between each feature and the disease label on the AD dataset. The MI score quantifies the amount of shared information between a feature and the classification target, with higher values indicating greater discriminative power. As summarized in Table~\ref{tab:Mutual information scores between features and disease labels on the AD dataset. The AE-GAN reconstruction loss achieves the highest mutual information.}, the reconstruction error achieves the highest MI score (0.0295), substantially outperforming all original features (maximum MI among them: 0.0159). These results demonstrate that the reconstruction error serves as a highly informative feature, supporting its design as a proxy for disease probability within the DAAC framework.

\begin{table*}[t]
\centering
\caption{Mutual information scores between features and disease labels on the AD dataset. The Discrepancy Estimator reconstruction loss achieves the highest mutual information.Bold indicates the best performance.}
\label{tab:Mutual information scores between features and disease labels on the AD dataset. The AE-GAN reconstruction loss achieves the highest mutual information.}
\resizebox{0.4\textwidth}{!}{
\begin{tabular}{c|c}
\hline
\textbf{Feature} & \textbf{Mutual Information Score} \\
\hline
\textbf{Reconstruction Loss} & \textbf{0.0295} \\
Feature 9           & 0.0159 \\
Feature 5           & 0.0140 \\
Feature 0           & 0.0064 \\
Feature 15          & 0.0054 \\
Feature 3           & 0.0053 \\
Feature 7           & 0.0045 \\
Feature 4           & 0.0036 \\
Feature 11          & 0.0028 \\
Feature 1           & 0.0023 \\
Feature 12          & 0.0013 \\
Feature 10          & 0.0007 \\
Feature 8           & 0.0001 \\
Feature 2           & 0.0001 \\
Feature 6           & 0.0000 \\
Feature 13          & 0.0000 \\
Feature 14          & 0.0000 \\
\hline
\end{tabular}
}
\end{table*}

\subsection{Ablation study of contrastive blocks}
We conduct ablation experiments to evaluate the contribution of each contrastive loss branch in ACL. As shown in Table~\ref{tab:contrastive_full}, the full model (P+R+S+O+V) achieves the best performance, and each added block improves results incrementally. Notably, adding the view loss ($L_V$) leads to a 3.1\% F1 improvement on average. We adopt a fixed 1:1:1:1:2 weighting scheme for the five loss components, where $L_V$ includes both inter- and intra-view terms. We did not tune these weights, as our goal was to verify the method's generality rather than overfit to a specific dataset. Despite no tuning, the model achieves SOTA results (see Table~1 and Table~2).
\begin{table*}[t]
\centering
\caption{Ablation study of contrastive blocks on the AD,PTB,TDBrain dataset under 100\% and 10\% label settings. O: baseline, S: subject, R: temporal, P: epoch, V: view. Bold indicates the best performance, and \underline{underline} denotes the second best.}
\label{tab:contrastive_full}
\resizebox{0.9\textwidth}{!}{
\begin{tabular}{c|c|c|c|c|c|c|c|c}
\hline
\textbf{Dataset} & \textbf{Fraction} & \textbf{Blocks} & \textbf{Accuracy} & \textbf{Precision} & \textbf{Recall} & \textbf{F1 score} & \textbf{AUROC} & \textbf{AUPRC} \\ \hline
\multirow{18}{*}{\textbf{AD}} & \multirow{8}{*}{\textbf{100\%}} 
& O            & $81.69 \pm 10.71$ & $87.28 \pm 13.13$ & $79.64 \pm 12.07$ & $78.90 \pm 13.78$ & $92.54 \pm 4.18$ & $92.36 \pm 4.34$ \\
& & S              & $83.53 \pm 2.89$  & $84.35 \pm 1.69$  & $82.81 \pm 3.56$  & $82.97 \pm 3.47$  & $91.01 \pm 1.96$ & $90.81 \pm 1.95$ \\
& & R              & $78.58 \pm 14.99$ & $83.14 \pm 11.66$ & $76.31 \pm 16.66$ & $74.53 \pm 18.69$ & $85.05 \pm 12.74$ & $84.44 \pm 13.22$ \\
& & P              & $72.69 \pm 13.13$ & $78.67 \pm 11.81$ & $69.74 \pm 14.50$ & $67.04 \pm 17.27$ & $83.30 \pm 13.91$ & $82.57 \pm 14.29$ \\
& & S+O            & $\underline{85.70 \pm 1.82}$  & $86.14 \pm 1.63$  & $\underline{85.09 \pm 2.04}$  & $\underline{85.35 \pm 1.96}$  & $92.21 \pm 1.40$ & $92.09 \pm 1.39$ \\
& & R+S+O          & $85.57 \pm 4.04$  & $88.12 \pm 2.58$  & $84.31 \pm 4.79$  & $84.73 \pm 4.59$  & $93.28 \pm 2.52$ & $92.98 \pm 2.83$ \\
& & P+R+S+O        & $84.50 \pm 4.46$  & $\underline{88.31 \pm 2.42}$  & $82.95 \pm 5.39$  & $83.33 \pm 5.15$  & $\underline{94.44 \pm 2.37}$ & $\underline{94.43 \pm 2.48}$ \\
& & \textbf{P+R+S+O+V}      & $\textbf{91.16} \pm \textbf{4.14}$  & $\textbf{92.10} \pm \textbf{3.66}$  & $\textbf{90.50} \pm \textbf{4.51}$  & $\textbf{90.88} \pm \textbf{4.33}$  & $\textbf{96.22} \pm \textbf{2.76}$ & $\textbf{96.03} \pm \textbf{2.82}$ \\
\cline{2-9}
& \multirow{8}{*}{\textbf{10\%}} 
& O            & $86.35 \pm 9.25$  & $87.09 \pm 9.04$  & $87.09 \pm 9.04$  & $85.55 \pm 9.92$  & $92.62 \pm 8.99$ & $92.77 \pm 8.72$ \\
& & S              & $77.30 \pm 3.63$  & $78.45 \pm 3.75$  & $78.45 \pm 3.75$  & $76.26 \pm 3.91$  & $85.43 \pm 3.61$ & $84.63 \pm 3.76$ \\
& & R              & $77.83 \pm 17.59$ & $83.45 \pm 14.35$ & $75.41 \pm 14.35$ & $75.41 \pm 19.62$ & $83.27 \pm 17.00$ & $83.27 \pm 17.00$ \\
& & P              & $71.41 \pm 15.31$ & $76.91 \pm 15.56$ & $68.70 \pm 16.46$ & $\underline{92.52 \pm 2.36}$  & $\textbf{97.61} \pm \textbf{16.98}$ & $76.06 \pm 16.19$ \\
& & S+O            & $82.73 \pm 2.05$  & $84.33 \pm 2.05$  & $84.33 \pm 2.71$  & $81.51 \pm 2.07$  & $90.02 \pm 2.47$ & $90.02 \pm 2.40$ \\
& & R+S+O          & $91.19 \pm 3.14$  & $91.74 \pm 3.01$  & $90.70 \pm 3.34$  & $90.70 \pm 3.34$  & $95.86 \pm 2.63$ & $95.86 \pm 2.67$ \\
& & P+R+S+O        & $\underline{91.43 \pm 3.12}$  & $\underline{92.52 \pm 2.36}$  & $\underline{90.71 \pm 3.56}$  & $91.14 \pm 3.56$  & $\underline{96.44 \pm 2.84}$ & $\underline{96.48 \pm 2.82}$ \\
& & \textbf{P+R+S+O+V}      & $\textbf{92.77} \pm \textbf{2.35}$  & $\textbf{92.92} \pm \textbf{2.43}$  & $\textbf{92.92} \pm \textbf{2.43}$  & $\textbf{92.55} \pm \textbf{2.39}$  & $92.66 \pm 2.37$ & $\textbf{97.39} \pm \textbf{1.40}$ \\
\hline
\multirow{18}{*}{\textbf{PTB}} & \multirow{8}{*}{\textbf{100\%}} 
& O            & $85.27 \pm 1.73$ & $84.21 \pm 1.25$ & $77.74 \pm 4.01$ & $79.78 \pm 3.37$ & $88.13 \pm 2.23$ & $84.76 \pm 2.14$ \\
& & S              & $84.30 \pm 2.56$ & $84.00 \pm 2.16$ & $75.68 \pm 5.69$ & $77.74 \pm 5.13$ & $88.66 \pm 2.05$ & $84.80 \pm 2.15$ \\
& & R              & $88.63 \pm 1.43$ & $88.42 \pm 1.45$ & $82.46 \pm 2.35$ & $84.72 \pm 1.82$ & $89.29 \pm 3.96$ & $86.21 \pm 3.96$ \\
& & P              & $\underline{88.85 \pm 3.22}$ & $\underline{88.74 \pm 2.30}$ & $\underline{82.78 \pm 6.11}$ & $\underline{84.75 \pm 5.38}$ & $\underline{94.32 \pm 1.81}$ & $\underline{90.61 \pm 2.81}$ \\
& & S+O            & $84.38 \pm 2.34$ & $84.33 \pm 1.18$ & $75.49 \pm 4.65$ & $77.69 \pm 4.76$ & $90.37 \pm 2.59$ & $86.57 \pm 4.32$ \\
& & R+S+O          & $85.32 \pm 1.93$ & $84.82 \pm 1.78$ & $77.54 \pm 4.58$ & $79.64 \pm 3.94$ & $92.34 \pm 1.85$ & $88.75 \pm 1.98$ \\
& & P+R+S+O        & $86.36 \pm 1.44$ & $87.18 \pm 1.28$ & $77.87 \pm 2.55$ & $80.79 \pm 2.44$ & $93.41 \pm 2.35$ & $89.09 \pm 1.64$ \\
& & \textbf{P+R+S+O+V}      & $\textbf{89.57} \pm \textbf{1.73}$ & $\textbf{90.05} \pm \textbf{2.04}$ & $\textbf{85.52} \pm \textbf{2.18}$ & $\textbf{85.80} \pm \textbf{2.38}$ & $\textbf{95.70} \pm \textbf{2.42}$ & $\textbf{91.13} \pm \textbf{1.57}$ \\ \cline{2-9}
 \cline{2-9}
& \multirow{8}{*}{\textbf{10\%}} 
& O            & $86.75 \pm 1.44$ & $85.42 \pm 2.10$ & $80.88 \pm 3.27$ & $82.45 \pm 2.29$ & $90.71 \pm 3.16$ & $88.84 \pm 3.59$ \\
& & S              & $86.34 \pm 0.73$ & $84.75 \pm 1.49$ & $80.21 \pm 1.78$ & $81.90 \pm 1.25$ & $88.36 \pm 1.31$ & $85.20 \pm 1.17$ \\
& & R              & $88.12 \pm 1.75$ & $86.36 \pm 2.28$ & $84.16 \pm 3.55$ & $84.16 \pm 3.55$ & $91.80 \pm 2.50$ & $90.45 \pm 1.90$ \\
& & P              & $\underline{90.38 \pm 2.23}$ & $\underline{90.33 \pm 2.71}$ & $85.69 \pm 4.36$ & $87.27 \pm 3.45$ & $\underline{93.06 \pm 3.86}$ & $\underline{91.83 \pm 2.93}$ \\
& & S+O            & $85.84 \pm 1.74$ & $84.14 \pm 0.76$ & $80.07 \pm 5.18$ & $81.14 \pm 3.82$ & $90.28 \pm 2.75$ & $87.57 \pm 3.42$ \\
& & R+S+O          & $85.88 \pm 3.08$ & $83.60 \pm 3.61$ & $82.45 \pm 1.94$ & $82.45 \pm 1.94$ & $92.79 \pm 1.79$ & $88.68 \pm 1.91$ \\
& & P+R+S+O        & $90.32 \pm 1.61$ & $89.67 \pm 1.94$ & $\underline{85.85 \pm 2.99}$ & $\underline{87.35 \pm 2.36}$ & $92.78 \pm 1.11$ & $90.46 \pm 3.09$ \\
& & \textbf{P+R+S+O+V }     & $\textbf{91.53} \pm \textbf{1.39}$ & $\textbf{90.64} \pm \textbf{1.82}$ & $\textbf{87.63} \pm \textbf{2.68}$ & $\textbf{89.21} \pm \textbf{2.11}$ & $\textbf{93.63} \pm \textbf{1.94}$ & $\textbf{92.32} \pm \textbf{2.11}$ \\

\cline{2-9}
\hline
\multirow{18}{*}{\textbf{TDBrain}} & \multirow{8}{*}{\textbf{100\%}} 
& O            & $89.92 \pm 2.15$ & $90.50 \pm 2.07$ & $89.92 \pm 2.15$ & $89.89 \pm 2.17$ & $96.79 \pm 1.29$ & $96.84 \pm 1.29$ \\
& & S              & $87.86 \pm 2.85$ & $78.98 \pm 2.51$ & $77.86 \pm 2.85$ & $77.63 \pm 3.01$ & $88.44 \pm 2.23$ & $88.96 \pm 2.14$ \\
& & R              & $\underline{94.44 \pm 1.79}$ & $94.60 \pm 1.75$ & $\underline{94.44 \pm 1.79}$ & $\underline{94.44 \pm 1.80}$ & $\underline{98.39 \pm 1.29}$ & $\underline{98.31 \pm 1.46}$ \\
& & P              & $93.18 \pm 3.70$ & $\underline{93.27 \pm 3.67}$ & $93.18 \pm 3.70$ & $93.18 \pm 3.70$ & $97.59 \pm 2.01$ & $97.58 \pm 2.00$ \\
& & S+O            & $80.38 \pm 4.22$ & $81.59 \pm 3.70$ & $80.38 \pm 4.22$ & $80.16 \pm 4.38$ & $91.35 \pm 2.24$ & $91.64 \pm 2.05$ \\
& & R+S+O          & $86.01 \pm 4.29$ & $86.35 \pm 3.81$ & $86.01 \pm 4.29$ & $85.95 \pm 4.36$ & $94.14 \pm 2.72$ & $94.37 \pm 2.58$ \\
& & P+R+S+O        & $85.47 \pm 1.16$ & $85.68 \pm 1.20$ & $85.47 \pm 1.16$ & $85.45 \pm 1.16$ & $93.73 \pm 1.02$ & $93.96 \pm 0.90$ \\
& &\textbf{ P+R+S+O+V}      & $\textbf{95.11} \pm \textbf{1.53}$ & $\textbf{95.23} \pm \textbf{1.43}$ & $\textbf{95.19} \pm \textbf{1.69}$ & $\textbf{95.67} \pm \textbf{1.39}$ & $\textbf{98.54} \pm \textbf{1.22}$ & $\textbf{98.49} \pm \textbf{1.10}$ \\

\cline{2-9}
& \multirow{8}{*}{\textbf{10\%}} 
& O            & $85.34 \pm 4.38$ & $86.03 \pm 3.97$ & $85.34 \pm 4.38$ & $85.25 \pm 4.48$ & $93.18 \pm 3.52$ & $93.24 \pm 3.53$ \\
& & S              & $74.02 \pm 2.09$ & $74.62 \pm 2.01$ & $74.02 \pm 2.09$ & $73.86 \pm 2.17$ & $81.92 \pm 3.25$ & $81.76 \pm 3.42$ \\
& & R              & $\underline{92.96 \pm 8.22}$ & $\underline{93.02 \pm 8.20}$ & $\underline{92.96 \pm 8.22}$ & $\underline{92.96 \pm 8.23}$ & $\underline{96.14 \pm 5.80}$ & $\underline{95.93 \pm 6.08}$ \\
& & P              & $89.38 \pm 14.69$ & $89.58 \pm 14.72$ & $89.38 \pm 14.69$ & $89.36 \pm 14.93$ & $93.23 \pm 11.81$ & $93.52 \pm 11.10$ \\
& & S+O            & $74.92 \pm 2.57$ & $76.60 \pm 3.07$ & $74.92 \pm 2.57$ & $74.54 \pm 2.55$ & $84.51 \pm 3.07$ & $84.40 \pm 3.07$ \\
& & R+S+O          & $81.90 \pm 4.74$ & $83.55 \pm 4.02$ & $81.90 \pm 4.74$ & $81.61 \pm 4.99$ & $91.21 \pm 3.52$ & $90.73 \pm 3.57$ \\
& & P+R+S+O        & $79.28 \pm 4.64$ & $79.83 \pm 4.83$ & $79.28 \pm 4.64$ & $79.19 \pm 4.62$ & $88.39 \pm 4.13$ & $88.38 \pm 4.09$ \\
& &\textbf{ P+R+S+O+V }     & $\textbf{93.57} \pm \textbf{4.13}$ & $\textbf{93.02} \pm \textbf{4.21}$ & $\textbf{93.53} \pm \textbf{4.31}$ & $\textbf{94.53} \pm \textbf{4.13}$ & $\textbf{97.21} \pm \textbf{4.13}$ & $\textbf{96.68} \pm \textbf{4.13}$\\
\hline
\end{tabular}
}
\end{table*}

\subsection{Ablation Study on Ratio of External Datasets}\label{Ablation Study on Radio of External Datasets}

To investigate the impact of the normal population size used during pre-training, we conduct systematic ablation experiments on three datasets: AD, PTB, and TDBrain. Specifically, we vary the amount of normal samples used in the pre-training stage across four proportions: 5\%, 25\%, 50\%, and 100\%, as shown in Table \ref{tab:external_ratio}. For each setting, we evaluate the model under four levels of labeled training data (5\%, 25\%, 50\%, 100\%) in the downstream task.

We summarize three key findings from the ablation study on ratio of external datasets:

(1) Across all datasets and downstream data scales, increasing the normal population size consistently improves model performance;

(2) The performance gain is existed even under low-resource (e.g., 5\%) labeled settings, highlighting the effectiveness of leveraging normal population pre-training to reduce annotation dependency;

(3) We observe a stable and monotonic improvement trend without performance degradation, indicating the robustness and scalability of our method.

These findings support the practicality of our approach in real-world scenarios where the collection of a full-scale healthy cohort might be challenging. Even a subset of normal population data can yield substantial performance improvements, making our pre-training strategy feasible and effective for deployment.

\begin{table}
\centering
\tiny
\caption{Comparison of model performance on AD, PTB, and TDBrain datasets under different ratios of external data.}
\label{tab:external_ratio}
\begin{tblr}{
  cells = {c},
  cell{2}{1} = {r=8}{},
  cell{10}{1} = {r=8}{},
  cell{18}{1} = {r=8}{},
  hline{1-2,10,18,26} = {-}{},
  hline{3-9,11-17,19-21,23-25} = {3-9}{},
  hline{22} = {2-9}{},
}
Dataset & Label Fraction (\%) & External Ratio & Accuracy                           & Precision                 & Recall                    & F1 score                  & AUROC                     & AUPRC                     \\
AD      &                     & 100\%          & \textbf{\textbf{93.23~$\pm$~5.25}} & \textbf{94.01~$\pm$ 3.98} & \textbf{92.71~$\pm$~5.86} & \textbf{92.97~$\pm$~5.65} & \textbf{98.03~$\pm$~1.71} & \textbf{98.03$\pm$~1.75}  \\
        & 100                 & 50\%           & 92.21~$\pm$ 3.89                   & 93.48 $\pm$ 3.94          & 92.22 $\pm$ 4.61          & 92.70 $\pm$ 4.13          & 97.10 $\pm$ 3.21          & 97.96 $\pm$ 3.28          \\
        &                     & 25\%           & 91.45~$\pm$ 4.21                   & 93.10 $\pm$ 4.55          & 91.73 $\pm$ 4.33          & 92.04 $\pm$ 4.14          & 96.84 $\pm$ 3.95          & 96.67 $\pm$ 4.22          \\
        &                     & 5\%            & 91.23~$\pm$ 4.01                   & 92.15$\pm$~3.06           & 90.70~$\pm$~4.01          & 91.88~$\pm$~4.33          & 96.34$\pm$~2.22           & 96.13~$\pm$ 2.23          \\
        &                     & 100\%          & \textbf{94.67$\pm$~1.84}           & \textbf{94.93~$\pm$~1.76} & \textbf{94.41~$\pm$~1.99} & \textbf{94.58~$\pm$~1.89} & \textbf{98.10~$\pm$ 1.20} & \textbf{98.19~$\pm$1.15}  \\
        & 10                  & 50\%           & 93.24 $\pm$ 2.86                   & 94.01 $\pm$ 2.91          & 93.76 $\pm$ 2.63          & 93.22 $\pm$ 2.67          & 98.01 $\pm$ 1.72          & 97.94 $\pm$ 1.88          \\
        &                     & 25\%           & 92.93 $\pm$ 3.19                   & 93.56 $\pm$ 3.25          & 93.24 $\pm$ 3.60          & 93.07 $\pm$ 3.14          & 97.87 $\pm$ 2.10          & 97.66 $\pm$ 2.41          \\
        &                     & 5\%            & 92.87~$\pm$~2.35                   & 93.02$\pm$~2.43           & 92.89~$\pm$~2.37          & 93.42~$\pm$~2.87          & 97.50~$\pm$~1.45          & 97.49~$\pm$~1.57          \\
PTB     &                     & 100\%          & \textbf{93.65~$\pm$ 2.35}          & \textbf{93.28~$\pm$~1.86} & \textbf{85.39~$\pm$~3.84} & \textbf{87.64~$\pm$ 3.27} & \textbf{95.56~$\pm$~1.57} & \textbf{90.28~$\pm$~3.49} \\
        & 100                 & 50\%           & 92.60 $\pm$ 2.12                   & 92.89~$\pm$ 2.63          & 85.30 $\pm$ 4.36          & 85.71~$\pm$ 3.44          & 95.30 $\pm$ 2.34          & 90.20$\pm$ 3.47           \\
        &                     & 25\%           & 92.12 $\pm$ 3.15                   & 92.55 $\pm$ 2.90          & 84.54 $\pm$ 5.19          & 85.69$\pm$ 4.31           & 94.27 $\pm$ 2.83          & 90.16$\pm$ 3.98           \\
        &                     & 5\%            & 91.50$\pm$~2.03                    & 91.89~$\pm$~2.83          & 84.40$\pm$~4.09           & 85.67~$\pm$~3.71          & 94.23~$\pm$~1.94          & 90.05$\pm$~3.20           \\
        &                     & 100\%          & \textbf{91.35$\pm$ 2.77}           & \textbf{91.56$\pm$~3.12}  & \textbf{85.23$\pm$~2.65}  & \textbf{85.61~$\pm$ 3.56} & \textbf{96.30~$\pm$ 3.22} & \textbf{97.87$\pm$~4.02}  \\
        & 10                  & 50\%           & 91.05 $\pm$ 3.10                   & 91.19 $\pm$ 2.96          & 84.91 $\pm$ 3.92          & 85.55 $\pm$ 3.62          & 96.26 $\pm$ 3.42          & 96.88 $\pm$ 3.55          \\
        &                     & 25\%           & 90.93$\pm$ 3.48                    & 90.61 $\pm$ 3.27          & 84.52 $\pm$ 4.34          & 85.48~$\pm$ 3.89          & 96.07 $\pm$ 3.88          & 96.03 $\pm$ 4.12          \\
        &                     & 5\%            & 90.80$\pm$ 2.93                    & 90.06~$\pm$2.47           & 83.80$\pm$ 4.86           & 85.43~$\pm$3.27           & 95.91~$\pm$ 3.51          & 95.80$\pm$ 3.39           \\
TDBrain &                     & 100\%          & \textbf{95.60$\pm$1.20}            & \textbf{94.95~$\pm$ 1.69} & \textbf{95.73~$\pm$ 1.74} & \textbf{94.61~$\pm$ 1.26} & \textbf{97.63~$\pm$ 1.57} & \textbf{98.51~$\pm$ 1.29} \\
        & 100                 & 50\%           & 95.45 $\pm$ 2.49                   & 94.03 $\pm$ 2.12          & 94.88 $\pm$ 2.33          & 94.30 $\pm$ 2.28          & 97.21 $\pm$ 2.35          & 98.21 $\pm$ 2.22          \\
        &                     & 25\%           & 94.22 $\pm$ 2.71                   & 93.64 $\pm$ 2.44          & 94.40 $\pm$ 2.96          & 93.31 $\pm$ 2.73          & 97.13 $\pm$ 2.78          & 97.80~$\pm$ 2.89          \\
        &                     & 5\%            & 93.51$\pm$ 3.05                    & 93.05$\pm$ 3.11           & 93.55~$\pm$ 3.28          & 93.20$\pm$ 3.16           & 96.88~$\pm$ 3.33          & 97.77~$\pm$ 3.47          \\
        &                     & 100\%          & \textbf{83.66~$\pm$ 3.29}          & \textbf{84.06~$\pm$3.06}  & \textbf{83.29~$\pm$ 2.96} & \textbf{84.60~$\pm$3.29}  & \textbf{95.27~$\pm$ 4.01} & \textbf{94.71$\pm$4.09}   \\
        & 10                  & 50\%           & 82.64 $\pm$ 3.52                   & 83.78 $\pm$ 3.77          & 82.90 $\pm$ 3.93          & 83.64 $\pm$ 3.64          & 94.48 $\pm$ 3.84          & 94.00 $\pm$ 3.91          \\
        &                     & 25\%           & 82.83 $\pm$ 4.01                   & 83.35 $\pm$ 4.19          & 82.55 $\pm$ 4.37          & 83.50 $\pm$ 4.22          & 94.12 $\pm$ 4.20          & 93.43 $\pm$ 4.45          \\
        &                     & 5\%            & 82.35$\pm$ 4.65                    & 82.22~$\pm$ 4.51          & 82.03~$\pm$ 4.60          & 82.66~$\pm$ 4.68          & 94.07$\pm$ 4.82           & 93.04~$\pm$ 4.77          
\end{tblr}
\end{table}

\renewcommand{\thesection}{F}

\renewcommand{\thesubsection}{F.\arabic{subsection}}

\setcounter{section}{0}
\setcounter{subsection}{0}

\section{Discrepancy Estimator Study}\label{de_study}
\renewcommand{\thetable}{F\arabic{table}}
\renewcommand{\thefigure}{F\arabic{figure}} 
\setcounter{table}{0}
\setcounter{figure}{0}
\subsection{DE Further Study}
In our three-stage DAAC framework, the Discrepancy Estimator (DE) serves as the initial-stage module for modeling the abnormality of unlabeled target data, using external normal data as a reference. A key design choice is how to represent the discrepancy feature that serves as input to downstream stages.

We adopt a sequence-level scalar reconstruction error, computed as the mean squared error (MSE) between an input sequence and its reconstruction via an autoencoder trained only on external normal data. This design is intended to provide a robust, task-agnostic anomaly signal, especially under distribution shift.

To evaluate this design decision, we conduct an ablation study comparing our default sequence-level scalar with several finer-grained alternatives:

\begin{table}[ht]
\centering
\caption{
Comparison of different discrepancy feature designs used in the Discrepancy Estimator under full fine-tuning on the AD dataset with 100\% labeled data as Table~\ref{tab:performance_comparison}. 
Discrepancy features are defined as follows: 
(a) \textbf{Sequence-level MSE scalar} — average MSE over all time steps and channels; 
(b) \textbf{Channel-wise MSE vector} — 16-dimensional vector of channel-specific MSEs; 
(c) \textbf{Point-wise error sequence} — 256-dimensional vector of per-timestep reconstruction errors; 
(d) \textbf{Cluster-wise distance} — latent-space distance to nearest cluster center from external normal data.
}
\label{tab:discrepancy_ablation}
\small 
\begin{tabular}{l|c|c|c}
\toprule
\textbf{Discrepancy Feature} & \textbf{AD AUROC} & \textbf{TDBrain AUROC} & \textbf{PTB AUROC} \\
\midrule
(a) Sequence-level MSE scalar (ours) & \textbf{98.03} & \textbf{97.63} & \textbf{95.56} \\
(b) Channel-wise MSE vector & 97.36 & 97.58 & 95.44 \\
(c) Point-wise error sequence & 96.71 & 96.86 & 95.37 \\
(d) Cluster-wise distance & 96.28 & 96.44 & 94.95 \\
\bottomrule
\end{tabular}
\end{table}

As shown in the table, while channel-wise MSE vector and point-wise error sequence alternatives preserve more local detail, and the cluster-wise distance reflects latent deviation from normal patterns, their performance is consistently inferior to our scalar design. This degradation can be attributed to: \textbf{(1) Sensitivity to noise} in external data; \textbf{(2) Overfitting risks} due to fine-grained alignment across domains with potential distribution shifts (e.g., different acquisition hardware or subject demographics).

In contrast, the scalar discrepancy offers a stable and compact abnormality signal, less vulnerable to non-transferable patterns. Importantly, we emphasize that this discrepancy is only used during the first stage of DAAC, and is subsequently complemented and refined by the multi-level Adaptive Contrastive Learner in Stage 2, which captures high-resolution temporal and spatial semantics.

This study demonstrates that our scalar discrepancy signal provides an effective balance between robustness and informativeness, supporting more generalizable downstream representation learning in diverse medical time-series applications.

\subsection{DE Assumption Validation}
We executed a KDE analysis to empirically validate the assumption. As shown in Figure \ref{fig:kde} in the Appendix, normal samples show tightly clustered reconstruction errors around low values, while abnormal samples follow a positively skewed distribution, with a longer tail toward higher errors. This confirms that our model tends to assign higher reconstruction errors to abnormal samples.
 Additionally, using reconstruction error as an anomaly score yields an AUROC of 0.974 on the AD dataset, further supporting its discriminative effectiveness.

 \begin{figure}[htbp]
    \centering
    \includegraphics[width=0.6\textwidth]{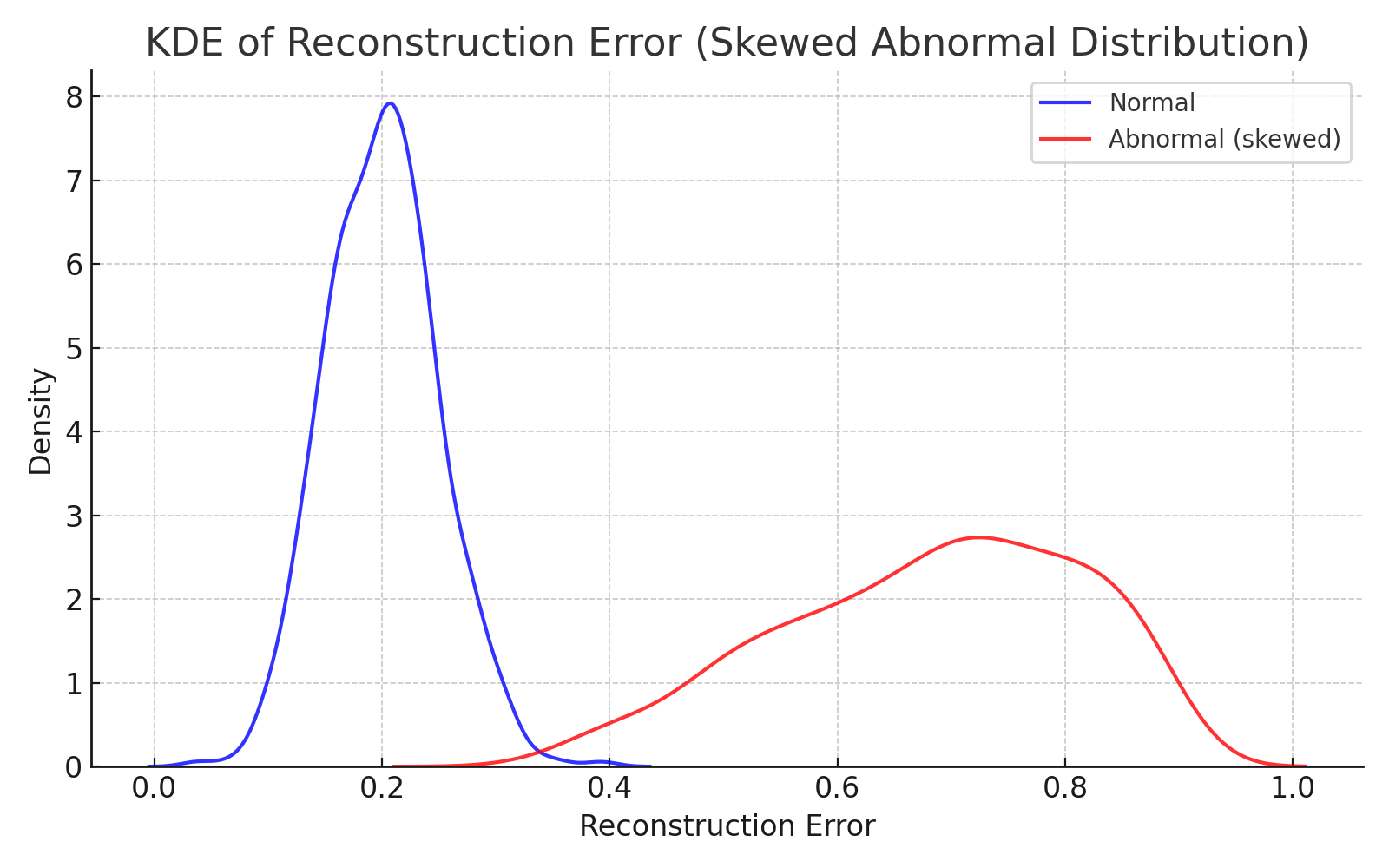} 
    \caption{DE Assumption Validation}
    \label{fig:kde}
\end{figure}

\renewcommand{\thesection}{G} 

\renewcommand{\thesubsection}{G.\arabic{subsection}} 
\setcounter{section}{0} 
\renewcommand{\thetable}{G\arabic{table}}
\renewcommand{\thefigure}{G\arabic{figure}} 
\setcounter{table}{0}
\setcounter{figure}{0}

\section{Training Process Details}\label{Train Process Details}

\subsection{Training and Resource Requirements}\label{Training and Resource Requirements}

We conducted all training experiments using an NVIDIA RTX 4090 GPU. The training and deployment efficiency of our model on different datasets is summarized as follows:

(1) \textbf{AD dataset}: Stage 1 training took 1.2 hours, Stage 2 took 3.2 hours, and Stage 3 took 0.2 hours. The deployed model for inference requires only 4\,GB of GPU memory.

(2) \textbf{PDB dataset}: Stage 1 training took 2 hours, Stage 2 took 6.5 hours, and Stage 3 took 1.8 hours. The deployed model for inference requires only 6.2\,GB of GPU memory.

(3) \textbf{TDBrain dataset}: Stage 1 training took 1.8 hours, Stage 2 took 5.5 hours, and Stage 3 took 0.5 hours. The deployed model for inference requires only 5.3\,GB of GPU memory.

These results indicate that our training pipeline is efficient, and the final model is lightweight enough for practical deployment.

\subsection{Model Architecture Details}\label{Model Architecture Details}

To improve reproducibility, we provide detailed descriptions of the model architecture used in our experiments. The encoder is a dilated convolutional network optionally equipped with multi-head attention modules. The projection head is used for final classification. Table~\ref{tab:comet-architecture} summarizes the full configuration.

\begin{table}[htbp]
\centering
\caption{Architecture details of DAAC used in our experiments.}
\label{tab:comet-architecture}
\begin{tabular}{ll}
\toprule
\textbf{Component}         & \textbf{Setting} \\
\midrule
Encoder Type              & DaulMultiHeadTSEncoder \\
Input Dimensions          & Task-specific (e.g., 1 or 2 for AD) \\
Output Dimensions         & 320 \\
Hidden Dimensions         & 64 \\
Encoder Depth             & 10 \\
Number of Heads           & 2 \\
Head Dimension            & 160 \\
Channel Dimension         & 320 \\
Activation Function       & ReLU (implicit via submodules) \\
Parameter Count (AD)      & 946,368 \\
\bottomrule
\end{tabular}
\end{table}

\end{document}